\documentclass[aps]{revtex4-1}
\usepackage[utf8]{inputenc}
\usepackage{amsmath}
\usepackage{amssymb}
\usepackage{geometry}
\usepackage{graphicx}
\usepackage{xcolor}
\usepackage{nicefrac}
\usepackage{hyperref}
 \usepackage[nottoc,numbib]{tocbibind}
 \newcommand{\figref}{Figure \ref}
 \newcommand{\Eqref}{Eq. \ref}
 \newcommand{\secref}{Section \ref}
 \newcommand{\xdownarrow}[1]{%
  {\left\downarrow\vbox to #1{}\right.\kern-\nulldelimiterspace}
}

\begin{document}

\title{Non-Abelian Geometric Phases Carried by the Spin Fluctuation Tensor}
\author{Bharath H. M.}
\date{\today}
\affiliation{School of Physics, Georgia Institute of Technology}

\begin{abstract}
The expectation values of the first and second moments of the quantum mechanical spin operator can be used to define a spin vector and spin fluctuation tensor, respectively. The former is a vector inside the unit ball in three space, while the latter is represented by an ellipsoid in three space. They are both experimentally accessible in many physical systems. By considering transport of the spin vector along loops in the unit ball it is shown that the spin fluctuation tensor picks up geometric phase information. For the physically important case of spin one, the geometric phase is formulated in terms of an SO(3) operator. Loops defined in the unit ball fall into two classes: those which do not pass through the origin and those which pass through the origin. The former class of loops subtend a well defined solid angle at the origin while the latter do not and the corresponding geometric phase is non-Abelian. To deal with both classes, a notion of generalized solid angle is introduced, which helps to clarify the interpretation of the geometric phase information. The experimental systems that can be used to observe this geometric phase are also discussed.

\end{abstract}
\maketitle


\section{Introduction}\label{sec 1}
Berry's geometric phase \cite{Berry.1984}, has attracted renewed interest in recent years due to applications in, for example, phase transitions with topological order parameters \cite{2016TopologicalPhasesWitten} and fault tolerant quantum computation \cite{2016GeometricQC, QUA:QUA24941}. Although Berry's phase was defined for adiabatic transport of a quantum system along a loop in the parameter space of the system's Hamiltonian, it was later realized that it is a kinematic property, which does not depend on the dynamics of the system \cite{Uhlmann.1986,PhysRevLett.58.1593, ANANDAN1988171}. In \cite{PhysRevLett.58.1593}, it is shown that a quantum system transported along a closed loops picks up a geometric phase irrespective of how the transport was induced. In \cite{Uhlmann.1986}, geometric phase has been generalized to open paths in the space of mixed states by defining an $SU(N)$ operator corresponding to every path in the space of $N\times N$ density matrices, with no reference to the question of how the transport along the path is induced. In other words, Berry's phase depends only on the path in the parameter space along which the quantum state is transported, and not on the dynamical equation governing the transport or the rate of transport. This insight has resulted in a kinematic formulation \cite{Mukunda-1}, \cite{Mukunda-2}, which we shall also adopt in this paper. The space of ground state eigenvectors of a non-degenerate Hamiltonian has a line bundle structure over its parameter space. Geometrically, Berry's phase can be viewed as the holonomy of Berry's connection form on this line bundle \cite{PhysRevLett.51.2167}. When the Hamiltonian is degenerate, Berry's phase generalizes to a non-Abelian Wilczek-Zee phase, which can also be formulated as a holonomy \cite{PhysRevLett.52.2111, ANANDAN1988171}. In general, geometric phase can be defined as a holonomy element of a connection form on a fiber bundle structure imposed on the space of quantum states \cite{chruscinski2004}, \cite{Avron1989}, \cite{1987JMP28.2102S}.

The geometric phase is essentially the geometric information stored in the overall phase of the wave-function of a quantum mechanical system. In this paper, we show that such geometric information may be extracted from second and higher order spin moments of a quantum spin system, which we formulate as a non-Abelian geometric phase. We restrict our analysis to pure quantum states, i.e., quantum states that can be represented by a vector in a finite-dimensional Hilbert space. Vectors in a finite dimensional Hilbert space may be regarded as states of a quantum spin system. Corresponding to every pure state, one can define a spin vector in real space as $\vec{s}= (\langle S_x\rangle, \langle S_y\rangle, \langle S_z\rangle )^T$, where $S_i$ are the Hilbert space spin operators and $\langle S_i\rangle$ are their expectation values with respect to the given pure state.  For a spin-$\frac{1}{2}$ system, the real space spin vector has unit length and therefore lies on the unit sphere; known as the Bloch sphere. For a spin-1 system, the length of the real space spin vector lies in the interval $[0,1]$ and therefore $\vec{s}$ lies in the closed unit ball, known as the Bloch ball ($\mathbb{B}$):
\begin{equation}
\mathbb{B} = \{ \vec{s} \in \mathbb{R}^3 : \ |\vec{s} | \leq 1\}
\end{equation}
A measure of the quantum fluctuations of the spin in the quantum state is given by the covariance matrix, a rank-2 tensor:
\begin{equation}\label{tensor}
\mathbf{T}=
\left(
\begin{array}{ccc}
\langle S_x^2\rangle-\langle S_x\rangle^2 & \frac{1}{2}\langle \{S_x,S_y\}\rangle -\langle S_x\rangle \langle S_y \rangle & \frac{1}{2}\langle \{S_x,S_z\}\rangle -\langle S_x\rangle \langle S_z \rangle\\
 \frac{1}{2}\langle \{S_x,S_y\}\rangle - \langle S_x\rangle \langle S_y \rangle &\langle S_y^2\rangle -\langle S_y\rangle^2 & \frac{1}{2}\langle \{S_z,S_y\}\rangle - \langle S_z\rangle \langle S_y \rangle\\
 \frac{1}{2}\langle \{S_x,S_z\}\rangle - \langle S_x\rangle \langle S_z \rangle & \frac{1}{2}\langle \{S_z,S_y\}\rangle -\langle S_z\rangle \langle S_y \rangle &\langle S_z^2\rangle - \langle S_z\rangle^2\\
\end{array}\right)
\end{equation}
Here, $\{S_i, S_j\}=S_iS_j + S_jS_i$ is the anticommutator of $S_i$ and $S_j$. Hereafter, we refer to this covariance matrix as the \textit{spin fluctuation tensor}.

\begin{figure}[h!]
\centering
\includegraphics[scale=0.7]{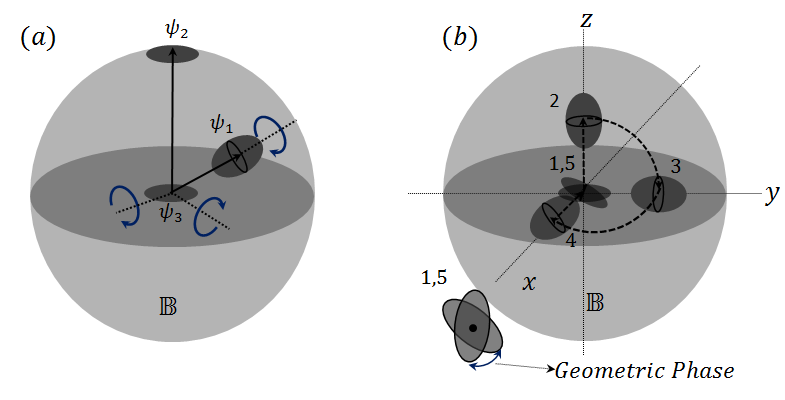}
\caption{\textbf{Spin vector and fluctuation tensor.} (a) Shows the Bloch ball $\mathbb{B}$ and three quantum states, $\psi_1, \psi_2 \ \text{and}\  \psi_3$, each one represented by its spin vector and the ellipsoid representing its spin fluctuation tensor (the ellipsoids are not to scale). $\psi_1$ has a spin vector with a length between $0 \ \text{and}\ 1$. The corresponding spin fluctuation tensor is represented by an ellipsoid. For a fixed spin vector, this ellipsoid has one degree of freedom (indicated in blue curved arrow). $\psi_2$ has a spin vector of unit length and the ellipsoid representing its spin fluctuation tensor is a disk.  $\psi_3$ is the zero spin vector and its spin fluctuation tensor is represented by a disk at the center. With the spin vector fixed to zero, this disk has two degrees of freedom (indicated by blue curved arrows). (b) Shows an example of a parallel transport of the ellipsoid along a loop and the resulting geometric phase. This loops starts and ends at the center of $\mathbb{B}$. In \secref{sec 4}, we show that the geometric phase of this loop is non-Abelian.}\label{FIG1}
\end{figure}

When the real space spin vector is transported along a loop in $\mathbb{B}$, the geometric phase information is encoded in the spin fluctuation tensor. To see this, we will introduce a geometric picture of the spin fluctuation tensor. The latter is a symmetric, positive semi-definite matrix with three non-negative eigenvalues and orthogonal eigenvectors. It may be represented by an ellipsoid whose principle axes have lengths given by the square-roots of the eigenvalues and whose orientation is determined by the eigenvectors. The pair $(\vec{s}, \mathbf{T})$ can be visualized by a vector in $\mathbb{B}$ with an ellipsoid representing the spin quantum fluctuations centered at its tip (\figref{FIG1}(a)). Let us consider a loop inside $\mathbb{B}$ along which the spin vector is transported. Analogous to the parallel transport of tangent vectors on a sphere, one can introduce a notion of parallel transport of the ellipsoids along the loop, where each of the ellipsoid's axes is parallel transported (Of course, the center of the Bloch ball poses a certain non-triviality which we address in this paper).  Upon circumscribing the loop the ellipsoid will, in general, return in a different orientation, capturing the geometric phase of the loop (\figref{FIG1}(b)). In this paper, we rigorously formulate this geometric phase as an element of the group $SO(3)$ and provide a geometric interpretation for the same.

The key step in formulating our geometric phase is to rigorously define parallel transport of the ellipsoids. In the sequel we show, however, that the parallel transport cannot be defined using the standard theory of connections on a fiber bundle. Hereafter, we restrict ourselves to spin-1 systems. The quantum state of a spin-1 system is represented by a non-zero vector $\psi=(z_{-1},z_0, z_{+1})^T$ in the 3-dimensional complex Hilbert space $\mathbb{C}^3$; here superscript $T$ denotes matrix transpose. The physical properties of the spin system are invariant under a multiplication of this vector by a non-zero complex scalar, i.e., $\psi$ and $\lambda \psi$ are physically equivalent state vectors for $\lambda \in \mathbb{C}-\{0\}$; this defines an equivalence class under the equivalence relation $ \psi \sim \lambda \psi$. The quotient space under this equivalence is a four dimensional manifold.  Topologically, the manifold is the \textit{complex projective plane} ($\mathbb{CP}^2$), defined as the space of all lines in a 3-dimensional complex vector space passing through the origin:
\begin{equation}
\mathbb{CP}^2 = \left\{   \psi \in \mathbb{C}^3-\{\vec{0}\}:   \ \psi\sim \lambda \psi\  \text{for}\  \lambda \in \mathbb{C}-\{0\} \right\}
\end{equation}
Equivalently, $\mathbb{CP}^2$ is the space of all 1-dimensional subspaces of $\mathbb{C}^3$. Each 1-dimensional subspace of $\mathbb{C}^3$ represents an equivalence class. The spin expectation values can be written as $\langle S_i\rangle = \frac{\langle \psi, S_i\psi\rangle}{\langle \psi, \psi\rangle}$, where $\langle \cdot , \cdot \rangle$ is the standard inner product on $\mathbb{C}^3$. We may define a map $\phi : \mathbb{CP}^2 \rightarrow \mathbb{B}$ that takes every equivalence class of $\mathbb{C}^3$ to its real space spin vector: $\phi(\psi)= \vec{s}$. In terms of coordinates, for a vector $\psi = (z_{-1}, z_0, z_{+1})^T$ representing the equivalence class $\{\lambda \psi: \lambda \in \mathbb{C}-\{0\}\}$, the image under this map is:
\begin{equation}\label{the map phi}
\phi\left(\left(\begin{array}{c}
z_{-1}\\
z_0\\
z_{+1}\\
\end{array}\right)\right)=\frac{1}{|z_{-1}|^2+|z_0|^2+|z_{+1}|^2} \left(\begin{array}{c}
\sqrt{2}\text{Re}(z_{-1}z_0^*+z_0z_{+1}^*)\\
\sqrt{2}\text{Im}(z_{-1}z_0^*+z_0z_{+1}^*)\\
|z_{+1}|^2-|z_{-1}|^2\\
\end{array}\right) = \vec{s} \in \mathbb{B}
\end{equation}
We note that the map $\phi$ is independent of the choice of the representative in any equivalence class and therefore, it is well defined. In other words, $\phi(\psi)=\phi(\lambda \psi)$ for $\lambda \in \mathbb{C}-\{0\}$. 

The components of the spin fluctuation tensor can also be written in terms of the coordinates of $\psi$. Together, the spin vector and the spin fluctuation tensor contain all the information about the spin-1 quantum state. Indeed, every spin-1 state is uniquely represented by the pair $(\vec{s}, \mathbf{T})$. Defining a parallel transport of ellipsoids along a loop in $\mathbb{B}$ is tantamount to defining a horizontal lift of loops in $\mathbb{B}$ to $\mathbb{CP}^2$. The map $\phi:\mathbb{CP}^2\rightarrow \mathbb{B}$ does not, however, have a fiber bundle structure. Any fiber bundle over $\mathbb{B}$ is necessarily a product bundle as $\mathbb{B}$ is a contractible space. The space $\mathbb{CP}^2$, being 4-dimensional, is not a product bundle over $\mathbb{B}$ because it has non-trivial second homology. Any 4-dimensional product bundle over $\mathbb{B}$, being homotopic to the 1-dimensional fiber itself, would have a trivial second homology. Therefore, this geometric phase cannot be formulated as a holonomy of loops in $\mathbb{B}$, in general. Circumventing this difficulty is the first of the two problems that we address in this paper.

\begin{figure}
\centering
\includegraphics[scale=0.7]{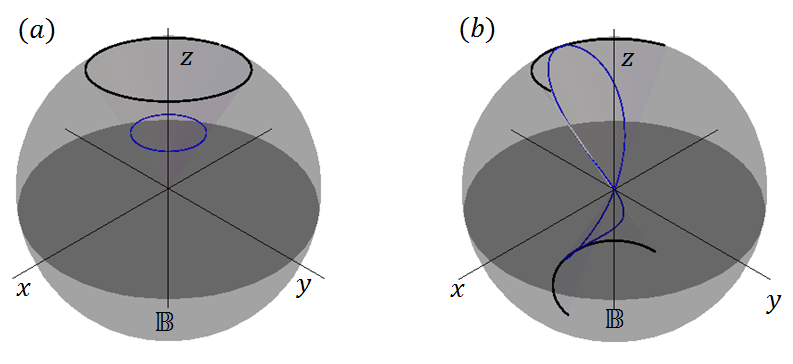}
\caption{\textbf{Non-singular and singular loops} (a) shows a non-singular loop inside the Bloch ball (in blue). Its solid angle is equal to the that of its radial projection to the surface (shown in black). (b) shows a singular loop inside the Bloch ball (in blue). This loop cannot be projected to the surface --- the center does not have a well-defined image under the projection. The curve in black is the projection of this loop excluding the point at the center. It does not have a well-defined solid angle. In \secref{sec 4}, we show that it has a well defined generalized solid angle and that it is equal to $\pi(2-\sqrt{3})$.}\label{FIG2}
\end{figure}

The interpretation of this geometric phase poses a separate problem. Berry's phase associated with a loop on the Bloch sphere, $\partial \mathbb{B}$, is generally interpreted as the solid angle enclosed by the loop \cite{Berry.1984}. The definition of solid angles easily extends to loops in the Bloch ball, provided they do not pass through the center. A convenient way to determine this solid angle is to radially project the loop onto the boundary $\partial \mathbb{B}$ where it subtends the same solid angle as the original loop (\figref{FIG2} (a)). We refer to such loops as \textit{non-singular} loops. Loops in $\mathbb{B}$ that pass through the center break into discontinuous pieces when projected to the boundary of $\mathbb{B}$; their solid angle cannot be defined by projection and therefore we refer to them as \textit{singular} loops (\figref{FIG2} (b)). It can be seen intuitively that for non-singular loops, the geometric phase is a rotation of the ellipsoid by an angle equal to the solid angle subtended by the loop, about the vector $\vec{s}$, at the base point of the loop. However, interpretation of the geometric phase of singular loops is non-trivial. It is the second problem we address in this paper. We restate these two problems as:

\begin{itemize}
\item[(i)] Given that $\mathbb{CP}^2$ is not a fiber bundle over $\mathbb{B}$, can we still define a curve in $\mathbb{CP}^2$ to be a horizontal lift of a loop in $\mathbb{B}$ and formulate a definition of geometric phase?
\item[(ii)] What is the interpretation of this geometric phase? In particular, can we attach a meaning to ``solid angles" for singular loops?
\end{itemize}
Pivotal to our solution of (i) is the idea that horizontal lifts in every known version of geometric phase minimize a certain metric in the fiber bundle \cite{Uhlmann1995}. In \secref{sec 2}A, we provide an outline of our solution to (i). As for the interpretation of geometric phases of singular loops, the term is justified by noting that the difficulty in defining solid angles for such a loop can not be solved by perturbing it and taking a limit \cite{Singular.Limits}. It requires a more detailed construction and a generalization of the notion of solid angles, which we provide in the next two sections and outline in \secref{sec 2}B. While \secref{sec 2} as a whole outlines all of our results, the details of definitions and proofs of the theorems therein are provided in \secref{sec 3}. Along with a few examples, we address the question of how to observe this geometric phase experimentally in \secref{sec 4}.

\section{Outline of Results}\label{sec 2}
 We state our solutions to (i) and (ii) in \secref{sec 2a} and \secref{sec 2b} respectively.
\subsection{Definition of Horizontal Lift And Geometric Phase}\label{sec 2a}

In definitions 1 and 2 below, we answer (i) by invoking the important role played by metrics in the theory of geometric phase \cite{Uhlmann1995}, \cite{Uhlmann2000}, \cite{Uhlmann2011}. In the definition of Berry's phase and Uhlmann's phase, horizontal lifts are constructed using Berry's connection form \cite{PhysRevLett.51.2167} and Uhlmann's connection form \cite{Uhlmann.1986}, respectively. It has been noted that in both of these cases, the horizontal lift can also be defined as the lift with minimal length in the respective fiber bundles \cite{Uhlmann1995, PATI1991105}. For a general Ehresmann connection \cite{Kobayashi}, if the horizontal subspace of the tangent space of a fiber bundle is defined as the orthogonal complement of the vertical subspace under a Riemannian metric, the resulting horizontal lift of a loop always minimizes the length among all lifts of the loop. While $\mathbb{CP}^2$ is not a fiber bundle over $\mathbb{B}$, it has a standard, natural (i.e., maximally symmetric) metric --- the Fubini-Study metric ($s_{FS}$) \cite{Geom.of.quant.states}. It is essentially the ``angle" between two quantum state vectors in the Hilbert space:
\begin{equation}\label{FS}
s_{FS}(\psi_1, \psi_2)= \cos^{-1} \left(\frac{|\langle \psi_1, \psi_2\rangle|}{\sqrt{\langle\psi_1, \psi_1 \rangle \langle \psi_2, \psi_2 \rangle}}\right)
\end{equation}
We note that this definition extends to $\mathbb{CP}^2,$ when we employ any Hilbert space representatives for the equivalence classes corresponding to the points in $\mathbb{CP}^2$, i.e., it is invaraint under scalar multiplications: $s_{FS}(\psi_1, \psi_2)=s_{FS}(\lambda_1\psi_1, \lambda_2\psi_2)$ where $\lambda_1, \lambda_2 \in \mathbb{C}-\{0\}$. We define a horizontal lift for loops in $\mathbb{B}$ using this metric.

\textbf{Definition 1 (Horizontal Lift):} A continuous path $\tilde{\gamma}:[0,1]\rightarrow \mathbb{CP}^2$  is called a \textit{horizontal lift} of a loop $\gamma:[0,1]\rightarrow \mathbb{B}$ iff $\phi\circ \tilde{\gamma} = \gamma$ and $\tilde{\gamma}$ minimizes the Fubini-Study length in $\mathbb{CP}^2$.

In \secref{sec 3}, we show that the earlier described intuitive notion of parallel transport of the ellipsoids along a loop in $\mathbb{B}$ is equivalent to the above definition of a horizontal lift of the loop. We show that, corresponding to every choice of $\tilde{\gamma}(0)$ satisfying $\phi(\tilde{\gamma}(0))=\gamma(0)$, there is a unique horizontal lift of $\gamma$.  In \Eqref{eqn_u}, we provide explicit equations to compute the horizontal lift of a given loop and a given initial point of the lift. Before proceeding to define a geometric phase using this horizontal lift, we note that not every loop in $\mathbb{B}$ has a well-defined horizontal lift in $\mathbb{CP}^2$. The relevant regularity conditions on the loop are summarized in theorem 1.

\textbf{Theorem 1 (Existence criteria for horizontal lifts):} A continuous, piece-wise differentiable loop $\gamma: [0,1]\rightarrow \mathbb{B}$ has a horizontal lift if it is differentiable at every $t\in [0,1]$ where $ \gamma(t)=\vec{0}\in\mathbb{B}$.

\begin{figure}[h!]
\centering
\includegraphics[scale=0.58]{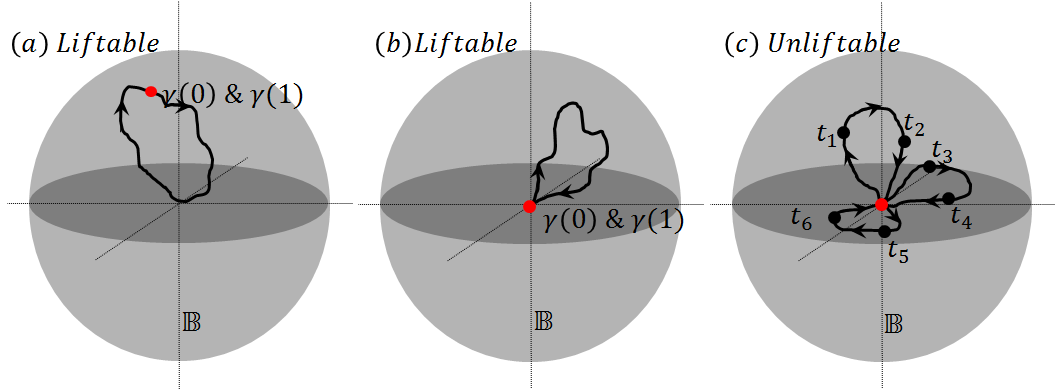}
\caption{\textbf{Liftable and unliftable loops.} (a) and (b) show liftable loops and (c) shows an unliftable loop. In all three loops, the red point represents the starting and the ending point (i.e., $\gamma(0)$ and $\gamma(1)$). For the loop in (a), $\gamma^{-1}(\vec{0})=\{t\}$ for some $t\in (0,1)$ and the loop is differentiable at that point. The loop in (b) has a kink at zero, but with a suitable choice of the starting an ending points, it is liftable. In particular, when the starting and the ending points are chosen at the center, i.e., $\gamma^{-1}(\vec{0})=\{0,1\}$, $\dot{\gamma}(0)$ and $\dot{\gamma}(1)$ are both well-defined and therefore, the loop is liftable. The loop in (c) has multiple kinks at the center. Six intermediate points between $0\ \& \ 1$, with the following ordering: $0<t_1<t_2<t_3<t_4<t_5<t_6<1$ are indicated to guide the reader through the loop.  There is no choice of the starting and the ending points such that it is liftable.  $\gamma^{-1}(\vec{0})$ has two points other than $0\ \&\ 1$ and the loop is not differentiable at either of these points. Therefore, this loop is not liftable. }\label{FIG3}
\end{figure}

This theorem essentially states that a loop in $\mathbb{B}$ has a horizontal lift if it has no ``kinks" while passing through the center of $\mathbb{B}$. We refer to the loops satisfying the condition mentioned in this theorem as \textit{liftable} loops. Clearly, any piece-wise differentiable loop not passing through the center of $\mathbb{B}$ is liftable. \figref{FIG3} shows two examples of liftable loops and one example of a loop that is not liftable. \figref{FIG3} (b) is an important example of a loop that appears to have a kink at the center of $\mathbb{B}$, but is nevertheless liftable. The apparent non-differentiability at the center is removable. If we choose the center as the starting and the ending points of the loop, i.e., $\gamma(0)=\gamma(1)=\vec{0}\in \mathbb{B}$, the loop satisfies all conditions mentioned in the theorem. However, the loop in \figref{FIG3} (c) is not liftable. There are multiple points of non-differentiability at the center, and so this loop does not satisfy the conditions mentioned in the above theorem. Therefore, a loop is liftable, if there is at least one parametrization under which it is differentiable at every visit to the center.

We now define geometric phase using the horizontal lift defined above. For a given loop $\gamma$ and a horizontal lift $\tilde{\gamma}$, the end points $\tilde{\gamma}(0)$ and $\tilde{\gamma}(1)$ are in $\mathbb{CP}^2$ and therefore, there is an operator $U\in SU(3)$ such that $\tilde{\gamma}(1)=U\tilde{\gamma}(0)$. This is because, $SU(3)$ acts transitively on $\mathbb{CP}^2$. The operator is not unique --- there are infinitely many such operators. Through its irreducible representation in $SU(3)$, $SO(3)$ can be regarded as a subgroup of $SU(3)$. We denote the representation as $\mathcal D : SO(3)\rightarrow SU(3)$. In \secref{sec 3}, we show that there is an $SO(3)$ choice for the operator $U$, i.e., there is an operator $R\in SO(3)$ with a representation $\mathcal{D}(R)\in SU(3)$ such that $\tilde{\gamma}(1)=\mathcal{D}(R)\tilde{\gamma}(0)$. However, this operator is still not unique --- it has a two fold ambiguity. We clear up this ambiguity and provide a more rigorous definition in \secref{sec 3}. We also show that this operator is independent of the choice of $\tilde{\gamma}(0)$, and so it is well-defined for $\gamma$. We define this $SO(3)$ operator as the geometric phase of $\gamma$.

\textbf{Definition 2 (Geometric Phase):} If $\gamma$ is a liftable loop in $\mathbb{B}$, its \textit{geometric phase} is the operator $R\in SO(3)$ such that, $\tilde{\gamma}(1)=\mathcal{D}(R)\tilde{\gamma}(0)$ holds for every horizontal lift $\tilde{\gamma}$ of $\gamma$, where $\mathcal{D}(R) \in SU(3)$ is the representation of $R$ in $SU(3)$.

In \secref{sec 3}, \Eqref{eqn_R}, we provide an explicit way of computing the geometric phase of a given loop. Going back to the earlier described geometric picture of representing a quantum state by a spin vector and an ellipsoid centered at its tip, the end points $\tilde{\gamma}(0)$ and $\tilde{\gamma}(1)$ are two quantum states with the same spin vector but different ellipsoids; i.e., we can represent them as $\tilde{\gamma}(0)\equiv(\vec{s}, \mathbf{T_1})$ and $\tilde{\gamma}(1)\equiv(\vec{s}, \mathbf{T_2})$. The geometric phase of $\gamma$, we show, is precisely the rotation $R$ which rotates the ellipsoid $\mathbf{T_1}$ to $\mathbf{T_2}$, i.e., $\mathbf{T_2} = R\mathbf{T_1}R^T$.

In the axis-angle representation, a right-hand rotation about a unit vector $\hat{n}\in \mathbb{R}^3$ by an angle $\theta \in [0,2\pi)$ is represented by $R_{\hat{n}}(\theta)$. For non-singular loops, the geometric phase is $R=R_{\gamma(0)}(\Omega)$, a rotation about the spin vector $\gamma(0)$ by an angle $\Omega$, equal to the solid angle enclosed by $\gamma$ (\figref{FIG1} (b)).  To see this, we need the following simple facts about the ellipsoids, which follow from \Eqref{the map phi}. One of the eigenvectors of $\mathbf{T}$ coincides with $\vec{s}$ with an eigenvalue $1-|\vec{s}|^2$. Therefore, the ellipsoid is always oriented with one axis parallel to $\vec{s}$ (see Appendix for a detailed derivation). The other two eigenvalues are $\frac{1}{2}(1\pm \sqrt{1-|\vec{s}|^2})$ and that leaves only one degree of freedom for the ellipsoid when the spin vector is fixed, namely, rotation about the spin vector (\figref{FIG1} (a)).  Therefore, if $\gamma(t)\neq 0$ throughout the loop, the geometric phase is necessarily a rotation about the vector $\gamma(0)$. The parallel transport of the ellipsoid is reminiscent of the parallel transport of a tangent line to $S^2$ along a loop and thus the holonomy is the solid angle of the loop. Therefore, the angle of rotation of the ellipsoid is also this solid angle.

The above interpretation, however, does not work for singular loops. We provide a generalization of the above interpretation in the following section.

\subsection{Interpretation of Geometric Phase}\label{sec 2b}

To answer point (ii), we define a \textit{generalized solid angle} for all loops inside $\mathbb{B}$ in definition 3 below.  The ideas is to first project a loop in the Bloch ball onto the \textit{real projective plane} ($\mathbb{RP}^2$) and to define a solid angle for this projection. We begin with the definition of the projection.

We recall that $\mathbb{RP}^2$ is the space of all lines through the origin of $\mathbb{R}^3$. Equivalently, it is the space obtained from the 2-sphere $S^2$ by identifying diametrically opposite points. We use the following notation for points in $\mathbb{RP}^2$:

\noindent \textbf{Notation:} The projection of a unit vector $\hat{n}\in S^2$, to $\mathbb{RP}^2$ is the equivalence class $\{+\hat{n}, -\hat{n}\}$ and will be denoted by $\pm \hat{n}$.

Every loop in $S^2$ can be projected to a loop in $\mathbb{RP}^2$.  As described earlier, the solid angle of a non-singular loop can be pictured by radially projecting it to the boundary of $\mathbb{B}$, which is $S^2$ (\figref{FIG2} (a)). A singular loop can also be projected to $S^2$ after removing the point(s)at the center. The projected path will, however, be discontinuous (\figref{FIG2} (b)). Every time the loop crosses the center of $\mathbb{B}$, the projected path makes a discontinuous jump across the diameter of $S^2$, parallel to the tangent of the loop at the center. This holds for all liftable loops. The discontinuity can be removed by identifying diametrically opposite points on $S^2$ and in doing so, we obtain an $\mathbb{RP}^2$. Thus, every liftable loop $\gamma$ in $\mathbb{B}$ can be projected to a continuous path $\alpha: [0,1] \rightarrow \mathbb{RP}^2$:
\begin{equation}\label{Projection to rp2}
\alpha (t) =
\begin{cases}
\pm \frac{\gamma (t)}{|\gamma (t)|} \qquad \gamma(t)\neq 0\\
\pm \frac{\dot{\gamma} (t)}{|\dot{\gamma} (t)|} \qquad \gamma(t)= 0\\
\end{cases}
\end{equation}
Here, $\dot{\gamma}=\frac{d\gamma}{dt}$. Note that the projection is in general an open path.

We will next define a solid angle for paths in $\mathbb{RP}^2$, as an appropriate $U(1)$ holonomy. Indeed the relevant $U(1)$ bundle over $\mathbb{RP}^2$ is isomorphic to the lens space $L(4,1)$. We recall that the lens space $L(4,1)$ is a quotient of the 3-sphere $S^3$ by the discrete group $Z_4$ action $(z_1,z_2) \mapsto (iz_1,iz_2)$, where $S^3$ is represented as the set of all normalized vectors in $\mathbb{C}^2$, i.e.,
\begin{equation}
S^3 = \{(z_1, z_2)\in \mathbb{C}^2: \ \  |z_1|^2+|z_2|^2=1\},
\end{equation}
and $Z_4= \{1, i, -1, -i\}$.
Thus $L(4,1)$ is obtained by identifying the orbits of $Z_4$ in $S^3$,
\begin{equation}\label{L41}
L(4,1)=\nicefrac{S^3}{(z_1,z_2)\sim (iz_1, iz_2)}
\end{equation}
$S^3$ is a 4-sheet covering space of $L(4,1)$.

The lens space $L(4,1)$ is a $U(1)$ bundle over both $\mathbb{RP}^2$ and $S^2$ (we will show this explicitly in \secref{sec 3}). In fact, this is the only lens space that is a $U(1)$ bundle over $\mathbb{RP}^2$ \cite{2016arXiv160806844G}. The solid angle of a loop in $S^2$ can be defined as the $U(1)$ holonomy of its lift in $L(4,1)$. Similarly, we define the solid angle of a loop in $\mathbb{RP}^2$ as the $U(1)$ holonomy of its lift in $L(4,1)$. An important property of this solid angle is that it is preserved under the projection map from $S^2$ to $\mathbb{RP}^2$ --- the solid angle of a loop in $S^2$ is equal to the solid angle of its projection in $\mathbb{RP}^2$. We prove this in lemma 3 in \secref{sec 3b}.  The appropriate generalization of a holonomy to open paths is a \textit{vertical displacement} \cite{Kobayashi}. The vertical displacement of the horizontal lift of a path in $\mathbb{RP}^2$ is a map from the fiber above the initial point of the path to the fiber above the final point of the path. Noting that $SO(3)\approx L(2,1)$ is a double cover of $L(4,1)$ and it acts transitively on $L(4,1)$, the vertical displacement can be represented by an $SO(3)$ action on $L(4,1)$, i.e., by an operator $V\in SO(3)$. We provide the details in \secref{sec 3}.

We now define the generalized solid angle of a loop in $\mathbb{B}$.

\textbf{Definition 3 (Generalized Solid Angle):} Let $\gamma$ be a liftable loop in $\mathbb{B}$ and let $\alpha$ be its projection in $\mathbb{RP}^2$. If $\tilde{\alpha}$ is a horizontal lift of $\alpha$ in $L(4,1)$ with a vertical displacement $V\in SO(3),$ and $\hat{k}$ is any unit vector normal to both $\alpha(0)$ and $\alpha(1)$, the generalized solid angle ($\Omega$) of the loop $\gamma$ is given by $\Omega = \cos^{-1}(\hat{k}\cdot V\hat{k})$.

In \secref{sec 3c}, we show that the expression $\Omega = \cos^{-1}(\hat{k}\cdot V\hat{k})$ is the correct holonomy of $\alpha$ when it is closed, and a meaningful definition of the solid angle of $\alpha$, also when it is open.  Furthermore, we also show that it is equal to the standard solid angle of $\gamma$ when it is non-singular. Hence we refer to this angle as the generalized solid angle of $\gamma$. The following theorem establishes the connection between the generalized solid angle and geometric phase:

\textbf{Theorem 2:} If $\gamma$ is a liftable loop in $\mathbb{B}$ and $\alpha$ is its projection in $\mathbb{RP}^2$, then the geometric phase of $\gamma$ is equal to the vertical displacement of $\alpha$.

Thus, the geometric phase of any loop inside $\mathbb{B}$ can be interpreted in terms of the generalized solid angle of its projection into $\mathbb{RP}^2$. This interpretation builds on the $m=0$ geometric phases introduced in \cite{Berry_m=0}. In the following section, we fill in the details of definitions 1, 2,  3  and provide proofs of theorem 1 and theorem 2. Before proceeding, we make a few remarks contrasting our geometric phase with Berry's phase. Unlike Berry's phase, our geometric phase does not arise naturally from the dynamics of the system. For any liftable loop inside $\mathbb{B}$, our geometric phase is well defined, regardless of how the physical system is transported along the loop. Therefore, our geometric phase is similar to the mixed state geometric phase introduced in \cite{Uhlmann.1986} and the non-adiabatic geometric phases introduced in
\cite{PhysRevLett.58.1593, ANANDAN1988171}. Both of these formulations have been observed experimentally \cite{EuroPhysLett.94.2.20007, PhysRevLett.60.1218}. In \secref{sec 4}, we briefly address the question of how to observe our geometric phase experimentally.

\section{Formulation And Proofs of Theorem 1 and Theorem 2}\label{sec 3}

The basic idea behind the proof of theorem 1 is that although $\phi: \mathbb{CP}^2\rightarrow \mathbb{B}$ does not have a fiber bundle structure, it is closely related to a fiber bundle. In fact, it can be constructed as a quotient of a fiber bundle. $\mathbb{B}$ can be constructed as a quotient space of $S^2\times[0,1]$, by collapsing the sphere $S^2\times\{0\}$ to a point.  We show in lemma 2(a) below that $\mathbb{CP}^2$ can also be constructed as a quotient space of $L(4,1)\times[0,1]$ by collapsing $L(4,1)\times \{0\}$ and $L(4,1)\times \{1\}$ to an $\mathbb{RP}^2$ and an $S^2$ respectively. $L(4,1)\times [0,1]$ is an $S^1$ bundle over $S^2\times[0,1]$, because $L(4,1)$ is an $S^1$ bundle over $S^2$. Thus, $\mathbb{CP}^2 \rightarrow \mathbb{B}$ can be constructed from the fiber bundle $L(4,1)\times [0,1]\rightarrow S^2 \times[0,1]$. Before proceeding to state and prove lemma 2, we develop a geometrical construction of $L(4,1)$. We show, in lemma 1, that $L(4,1)$ is the space of all tangent lines to a unit sphere.

\textbf{Lemma 1:} $L(4,1)$ is homeomorphic to the space of all tangent lines to a unit sphere and it is an $S^1$ bundle over both $S^2$ and $\mathbb{RP}^2$.

\begin{figure}[h!]
\centering
\includegraphics[scale=0.7]{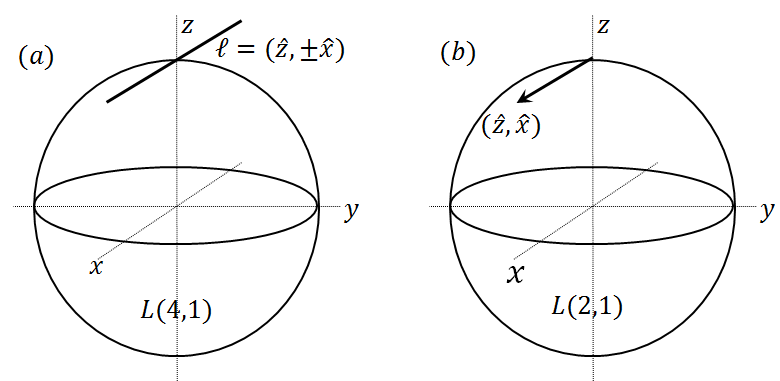}
\caption{\textbf{The lens spaces $L(4,1)$ and $L(2,1)$.} $L(4,1)$ is the space of all tangent lines to a sphere and $L(2,1)$ is the space of all unit tangent vectors to a sphere. (a) shows the tangent line $\ell = (\hat{z},\pm \hat{x}) \in L(4,1)$, parallel to $\hat{x}$ and touching the sphere at $\hat{z}$. (b) shows a unit tangent vector to a sphere $(\hat{z}, \hat{x})\in L(2,1)$ at $\hat{z}$ parallel to $\hat{x}$.}\label{FIG4}
\end{figure}

\textbf{Proof:} A tangent line ($\ell$) to a sphere is uniquely represented by the pair $\ell = (\hat{v}, \pm \hat{u})$ (\figref{FIG4}(a)) of orthogonal unit vectors, $\hat{v}$ representing the point of tangency of $\ell$ and $\hat{u}$ representing the direction of $\ell$. Here, $-\hat{u}$ and $+\hat{u}$ represent the same tangent line and therefore, we use a $``\pm"$ sign before $\hat{u}$, as a short hand for the equivalence class $\{+\hat{u}, -\hat{u}\}$. We show that the space of all tangent lines to a sphere, i.e, $\{\ell = (\hat{v}, \pm \hat{u}): \quad \hat{v}\cdot \hat{u}=0\}$ is homeomorphic to $L(4,1)$ by explicitly constructing a 4-sheeted covering map from $S^3$ to this space and showing that this space is also obtained as a quotient of $S^3$ under a $Z_4$ action (\Eqref{L41}).

Noting that $SU(2)$ is topologically homeomorphic to $S^3$ and $SO(3)$ acts transitively on the space of tangent lines to a sphere, we construct a composition of the following two maps:
\begin{equation}
SU(2)\xrightarrow[]{f} SO(3) \xrightarrow[]{g} \{\ell = (\hat{v}, \pm \hat{u}): \quad \hat{v}\cdot \hat{u}=0\}
\end{equation}
 $f$ is the standard double cover from $SU(2)$ to $SO(3)$ i.e., $f : e^{i\hat{n}\cdot \vec{\sigma} \frac{\theta}{2}} \mapsto R_{\hat{n}}(\theta)\in SO(3) $, where $\hat{n}$ is a unit vector in $\mathbb{R}^3$ and  $\vec{\sigma}=(\sigma_x, \sigma_y, \sigma_z)$ are the Pauli matrices:
\begin{equation}
\sigma_x=\left(\begin{array}{cc}
0 & 1\\
1 & 0\\
\end{array}\right), \quad \sigma_y=\left(\begin{array}{cc}
0 & -i\\
i & 0\\
\end{array}\right), \quad \sigma_z=\left(\begin{array}{cc}
1 & 0\\
0 & -1\\
\end{array}\right)
\end{equation}
The map $g$ is constructed from the action of $SO(3)$ on the space of tangent lines to a sphere. Fixing a tangent line $\ell_0 = (\hat{z}, \pm \hat{x})$ (\figref{FIG4}(a)), we obtain:
\begin{equation}
g : R_{\hat{n}}(\theta) \mapsto R_{\hat{n}}(\theta)\ell_0 = \left(R_{\hat{n}}(\theta)\hat{z}, \pm R_{\hat{n}}(\theta)\hat{x}\right)
\end{equation}
We now show that $g \circ f : SU(2)\rightarrow \{(\hat{v}, \pm\hat{u}):\quad \hat{v}\cdot \hat{u}=0\}$ is the required 4-sheet covering map. The action of $SO(3)$ on a tangent line to a sphere has a $Z_2$ stabilizer. For instance, the stabilizer of $\ell_0$ is $\{1,R_{\hat{z}}(\pi)\}$. Therefore, $g$ is a double covering map. For an arbitrary tangent line $\ell$, the pre-image set under $g$ contains two points in $SO(3)$. If $\ell = R_{\hat{n}}(\theta)\ell_0$, for some $\hat{n}$ and $\theta$, then its pre-image set is $g^{-1}(\ell)=\{R_{\hat{n}}(\theta), R_{\hat{n}}(\theta)R_{\hat{z}}(\pi)\}$. Further, $f^{-1}\circ g^{-1}(\ell)$ is a set of 4 elements in $SU(2)$ given by:
\begin{equation}
f^{-1}\circ g^{-1}(\ell)= e^{i\hat{n}\cdot \vec{\sigma} \frac{\theta}{2}}\{1, i\sigma_z, -1, -i\sigma_z\}
\end{equation}
Thus, the pre-image set is generated by a $Z_4$ action and therefore, $g\circ f$ is the required covering map and $L(4,1)\approx \{\ell = (\hat{v}, \pm \hat{u}): \quad \hat{v}\cdot \hat{u}=0\}$. We can now define the bundle maps $\pi_1 : L(4,1)\rightarrow S^2$ and $\pi_2 : L(4,1)\rightarrow \mathbb{RP}^2$:
\begin{equation}
\begin{split}
\pi_1: (\hat{v}, \pm \hat{u}) &\mapsto \hat{v}\in S^2\\
\pi_2 :  (\hat{v}, \pm \hat{u})&\mapsto \pm\hat{u}\in \mathbb{RP}^2\\
\end{split}
\end{equation}
$\pi_1$ takes every tangent line to its point of tangency, and $\pi_2$ takes every tangent line to a parallel line through the center, which is an element of $\mathbb{RP}^2$. It is straight forward  to verify that they are both $S^1$ bundle maps $\blacksquare$.

A natural metric on $L(4,1)$ is induced by the round metric (i.e., the standard Cartesian metric) on $S^3$. This metric, at a point $\ell = (\hat{v}, \pm \hat{u})\in L(4,1)$ is:
\begin{equation}\label{Metric on L41}
ds^2 = d\hat{v}\cdot d\hat{v} + d\hat{u}\cdot d\hat{u}-(\hat{v}\cdot d\hat{u})^2
\end{equation}
The first term ($d\hat{v}\cdot d\hat{v}$) corresponds to the distance covered by the point of contact on $S^2$. The term $d\hat{u}\cdot d\hat{u}-(\hat{v}\cdot d\hat{u})^2$ corresponds to the angle of rotation of the tangent line about its point of contact.

Using a similar argument, it can be shown that the lens space $L(2,1)$ is the space of all unit tangent vectors to a unit sphere, i.e.,  $L(2,1) \approx \{(\hat{v}, \hat{u}):\quad \hat{u}\cdot \hat{v}=0\}$ (\figref{FIG4} (b)).

\textbf{Lemma 2:}
\begin{itemize}
\item[(a)] $\mathbb{CP}^2$ can be constructed from the stack $L(4,1)\times [0,1]$ by collapsing $L(4,1)\times\{0\}$ to an $\mathbb{RP}^2$ and $L(4,1)\times\{1\}$ to an $S^2$ using the respective bundle maps $\pi_1$ and $\pi_2$. That is,
\begin{equation}
\mathbb{CP}^2 = \nicefrac{L(4,1)\times [0,1]}{\pi}
\end{equation}
where $\pi=1$ on $L(4,1)\times(0,1)$, $\pi=\pi_1$ on $L(4,1)\times\{1\}$ and  $\pi=\pi_2$ on $L(4,1)\times\{0\}$
\item[(b)] Writing $\mathbb{B}^{\circ}-\{0\} = S^2 \times(0,1)$, where $\mathbb{B}^{\circ}$ is the interior of $\mathbb{B}$, the restriction of $\phi$ to $L(4,1)\times(0,1)$ is,
\begin{equation}
\phi = \pi_1\times 1 : L(4,1)\times (0,1) \rightarrow S^2\times (0,1).
\end{equation}
\item[(c)] $\mathbb{CP}^2$ is the space of all chords to a unit sphere and $\phi$ maps each chord to its center.
\end{itemize}

\textbf{Proof:}
We begin with a proof of $(a)$. Let us consider the pre-image sets of $\phi$:
\begin{equation}\label{pre-image sets}
\phi^{-1}(\vec{s})=\begin{cases}
\mathbb{RP}^0 \text{ if } |\vec{s}|=1 \\
\mathbb{RP}^1 \text{ if } 0<|\vec{s}|<1 \\
\mathbb{RP}^2 \text{ if } |\vec{s}|=0 \\
\end{cases}
\end{equation}
This can be shown using the explicit algebraic expression of $\phi$,
\Eqref{the map phi}. However, it is more illuminating to use the earlier described geometric picture of representing a point in $\mathbb{CP}^2$ as a vector and an ellipsoid, i.e., $(\vec{s}, \mathbf{T})$ (\figref{FIG1} (a)). The lengths of the axes of the ellipsoid are $1-|\vec{s}|^2$, $\frac{1}{2}(1\pm \sqrt{1-|\vec{s}|^2})$ (see Appendix). Therefore, its dimensions depend only on the length of the spin vector. When $|\vec{s}|\neq 0$, one of its axes is parallel to $\vec{s}$. For a given spin vector with $0<|\vec{s}|<1$, the ellipsoid has one degree of freedom --- rotation about $\vec{s}$, which produces the set of all quantum states with spin vector $\vec{s}$. This set is an $\mathbb{RP}^1$, because the ellipsoid has a two fold symmetry when rotated about $\vec{s}$.

On the boundary of $\mathbb{B}$, when $|\vec{s}|=1$, the lengths of the two transverse axes of the ellipsoid are equal and the length of the third axis is zero. Therefore, the ellipsoid degenerates into a disk perpendicular to $\vec{s}$. It has no degrees of freedom; it is the only quantum state with the given spin vector. Thus, the pre-image set of this spin vector is just a point i.e., $\mathbb{RP}^0$.

Finally, when $|\vec{s}|=0$, the ellipsoid again degenerates to a disk at the center of $\mathbb{B}$. This time, however, it has two degrees of freedom. The pre-image set $\phi^{-1}(\vec{0})$ is the space of all orientations of a disk in $\mathbb{R}^3$ centered at the origin. This is indeed $\mathbb{RP}^2$.

It follows, now, that the pre-image set of the boundary of $\mathbb{B} $, i.e., $\phi^{-1}(\{\vec{s}: \quad |\vec{s}|=1\})$ is a sphere in $\mathbb{CP}^2$. For a shell of radius $0<r<1$, the pre-image set is a lens space $L(4,1)$:
\begin{equation}
\phi^{-1}(\{\vec{s}: \quad |\vec{s}|=r\})= L(4,1) \quad 0<r<1
\end{equation}
To show this, we use lemma 1 and construct a bijective map from the pre-image of the shell to $L(4,1)$. Consider the map $(\vec{s}, \mathbf{T})\mapsto (\hat{v}, \pm \hat{u})$ where $\hat{v}=\frac{\vec{s}}{r}$ and $\hat{u}$ is the eigenvector of $\mathbf{T}$ normal to $\vec{s}$, with the smaller eigenvalue. Indeed, there is a one-one correspondence between the orientations of an ellipsoid at $\vec{s}$ and tangent lines at $\vec{s}$ to a sphere of radius $|\vec{s}|$. Thus, it follows from lemma 1 that the pre-image of a shell is homeomorphic to $L(4,1)$.

We can now construct $\mathbb{CP}^2$ using the pre-image sets:
\begin{equation}
\begin{split}
\phi^{-1}(\{\vec{s}: \quad |\vec{s}|=1\})&= S^2\\
\phi^{-1}(\{\vec{s}: \quad 0<|\vec{s}|<1\})&= L(4,1)\times(0,1)\\
\phi^{-1}(\vec{0})&= \mathbb{RP}^2\\
\end{split}
\end{equation}
$\mathbb{CP}^2$ is therefore obtained by attaching an $\mathbb{RP}^2$ and an $S^2$ to either ends of $L(4,1)\times(0,1)$. The attaching maps are easily seen to be $\pi_1$ and $\pi_2$, using the geometric picture. Thus, $\mathbb{CP}^2$ is obtained from $L(4,1)\times [0,1]$ by collapsing $L(4,1)\times\{0\}$ to an $\mathbb{RP}^2$ and $L(4,1)\times\{1\}$ to an $S^2$ using the respective bundle maps.

$(b)$ follows trivially from the above construction of pre-image sets. The geometrical construction of $\mathbb{CP}^2$ claimed in $(c)$ can be shown as follows. The chords passing through the center of a unit sphere form an $\mathbb{RP}^2$. The chords at some distance $r\in(0,1)$ from the center form an $L(4,1)$ and the chords at a distance $1$ from the center degenerate to points on a sphere, forming a sphere. Thus, the space of all chords to a unit sphere has the same structure as $\mathbb{CP}^2$ and is homemorphic to it. $\blacksquare$.

Lemma 2(c) is also a consequence of Majorana constellation \cite{Majorana1932} which has been used very fruitfully to understand geometric phases \cite{PhysRevLett.108.240402}. States of a spin-1 system can be considered as symmetric states of a two coupled spin-$1/2$ systems. A spin-$1/2$ state is a point on a Bloch sphere (i.e., $\mathbb{CP}^1$) and therefore, a spin-1 state is an unordered pair of points on the Bloch sphere (see ref. \cite{Geom.of.quant.states} for a detailed description of this representation). This is equivalent to a chord\footnote{This picture has a generalization. $\mathbb{CP}^n$ is a \textit{unordered product} of $n$ $\mathbb{CP}^1$'s. It is the space of all unordered set of $n$ points on a unit sphere. That is,  $\mathbb{CP}^n = \mathbb{CP}^1\times \cdots \times \mathbb{CP}^1/\sim$ where $(r_1,\cdots r_i,\cdots r_j,\cdots r_n)\sim(r_1,\cdots r_j,\cdots r_i,\cdots r_n)$ for $r_i \in \mathbb{CP}^1$. This is known as Majorana constellation.}. $\phi$ maps each chord to its center.

We can represent a chord as $(r, \hat{v}, \pm \hat{u})$, where $r\hat{v}$ is the center of the chord and $\hat{u}$ is its direction. This corresponds to a quantum state whose spin vector is $r\hat{v}$ and the ellipsoid is oriented such that the eigenvector normal to $\vec{s}$ with the smaller eigenvalue is parallel to $\hat{u}$. It is straightforward to construct this quantum state $\psi \in \mathbb{C}^3$. For instance, written in the standard basis,
 \begin{equation}\label{CP2-chord}
 (r, \hat{z}, \pm \hat{x})\mapsto \psi = \left(\sqrt{\frac{1-r}{2}}, 0, \sqrt{\frac{1+r}{2}}\right) \in \mathbb{C}^3
 \end{equation}
Quantum states corresponding to any chord can be obtained by preforming rotations on both sides of the above equation. Conversely, the chord corresponding to given quantum state can be obtained from its spin vector and fluctuation tensor, $(\vec{s}, \mathbf{T})$ --- it is the chord centered at $\vec{s}$ and oriented parallel to the largest axis of $\mathbf{T}$ perpendicular to $\vec{s}$.

The Fubini-Study metric on $\mathbb{CP}^2$ can be applied to the space of all chords to a unit sphere.  At $(r, \hat{v}, \pm \hat{u})$, the metric is:
\begin{equation}\label{Fubini-Study}
ds_{FS}^2 = \frac{1}{2}(1-\sqrt{1-r^2})d\hat{v}\cdot d\hat{v}+ \sqrt{1-r^2}(\hat{u}\cdot d\hat{v})^2+ (1-r^2)(d\hat{u}\cdot d\hat{u}-(\hat{v}\cdot d\hat{u})^2)+\frac{1}{4(1-r^2)}dr^2
\end{equation}
This follows from \Eqref{FS}. We now proceed to prove theorem 1.

\subsection{Proof of theorem 1}\label{sec 3a}
Without loss of generality, we may assume that $\dot{\gamma}(t)\neq 0$ whenever it is well-defined. Therefore, $\gamma^{-1}(\vec{0})$ is a zero dimensional compact manifold, i.e., it is a finite set of points. Adding the end points $0 \ \text{and}\ 1$ to this finite set, we obtain a set of points,  $\gamma^{-1}(\vec{0})\cup \{0,1\}=\{a_0,\cdots a_{n+1}\}$ where, $a_i<a_{i+1}$, $a_0=0$ and $a_{n+1}=1$. This set divides the loop into $n+1$ pieces, $\gamma_{j}:[a_{j-1},a_j]\rightarrow \mathbb{B}$ for $j=1,2,\cdots n+1 $. Each piece $\gamma_j$ may start and end at the center of $\mathbb{B}$, but lies away from the center otherwise. That is, its interior lies away from the center, $\gamma((a_{j-1},a_j))\subset S^2\times(0,1]$. The closure of this path in $S^2\times [0,1]$ has a horizontal lift in $L(4,1)\times [0,1]$, defined using the standard theory of connections \cite{Kobayashi},  because this space has a circle bundle structure over $S^2\times[0,1]$. We denote this horizontal lift by $ \tilde{\gamma}_j: [a_{j-1},a_j]\rightarrow L(4,1)\times[0,1]$. This path can be projected to $\mathbb{CP}^2$ by composing it with $\pi$, as shown in lemma 2(a). The idea behind this proof is to show that these  projected paths can be attached continuously under the assumptions of the theorem, and the resulting path in $\mathbb{CP}^2$ is a lift of $\gamma$ that minimizes the Fubini-Study length.

Within $(a_{j-1}, a_j)$, we may write $\gamma_j(t) = (\frac{\gamma_j(t)}{|\gamma_j(t)|}, |\gamma_j(t)|)\in S^2\times(0,1]$ where the two components represent the coordinates in $S^2$ and $(0,1]$ respectively, i.e., $\frac{\gamma_j(t)}{|\gamma_j(t)|}\in S^2$ and $|\gamma_j(t)|\in (0,1]$. Let us define the closure of the first component as $\beta_j: [a_{j-1},a_j]\rightarrow S^2$:
\begin{equation}\label{beta}
\beta_j(t) = \begin{cases}
\frac{\gamma_j(t)}{|\gamma_j(t)|} \quad a_{j-1}<t<a_j\\
\lim_{t'\rightarrow a_k} \frac{\gamma_j(t')}{|\gamma_j(t')|} \quad t=a_k,  \ k =j, j-1\\
\end{cases}
\end{equation}
Note that $\beta_j$ are indeed the closures of the discontinuous radial projections shown in \figref{FIG2} (b). Let $\tilde{\beta}_j$ denote a horizontal lift of $\beta_j$ in $L(4,1)$. We define paths $\tilde{\gamma}_j:[a_{j-1},a_j]\rightarrow L(4,1)\times[0,1]$ as:
 \begin{equation}
 \tilde{\gamma}_j(t)= (\tilde{\beta}_j(t), |\gamma_j(t)|)
\end{equation}
We next show that after projecting these paths to $\mathbb{CP}^2$, i.e., $\pi\circ \tilde{\gamma}_j$ can be attached continuously at all $a_j$ for $j=1, 2\cdots n$. Note that $\gamma(a_j)=\vec{0}$ for $j=1,2\cdots n$. The end points of the two neighboring paths, $\tilde{\gamma}_j$ and $\tilde{\gamma}_{j+1}$ at $a_j$, projected to $\mathbb{CP}^2$ are given by:
\begin{equation}
\begin{split}
\pi \circ \tilde{\gamma}_j(a_{j})=\pi \circ (\tilde{\beta}_j(a_j), 0) &\equiv \pi_2\circ \tilde{\beta}_j(a_j)\in \mathbb{RP}^2 = \phi^{-1}(\vec{0})\\
\pi \circ \tilde{\gamma}_{j+1}(a_{j})=\pi \circ (\tilde{\beta}_{j+1}(a_j), 0) &\equiv \pi_2\circ \tilde{\beta}_{j+1}(a_j)\in \mathbb{RP}^2 =\phi^{-1}(\vec{0})\\
\end{split}
\end{equation}
It suffices to show that the first point of the lift, $\tilde{\beta}_{j+1}(a_j)$, can be chosen such that the above two points coincide in $\mathbb{CP}^2$. We begin with a simple observation; since $\gamma$ is liftable, it is differentiable at $a_j$ and it follows that \footnote{We have used $ \lim_{t\rightarrow a_k^{\pm}} \frac{\gamma_j(t)}{|\gamma_j(t)|} =  \lim_{t\rightarrow a_k^{\pm}} \frac{\gamma_j(t)-\gamma_j(a_k)}{|\gamma_j(t)-\gamma_j(a_k)|}= \pm\frac{\dot{\gamma}_j(a_k)}{|\dot{\gamma}_j(a_k)|}$}:
\begin{equation}
\begin{split}
\beta_j(a_j) = \lim_{t\rightarrow a_j}\frac{\gamma_j(t)}{|\gamma_j(t)|} = \frac{\dot{\gamma}(a_j)}{|\dot{\gamma}(a_j)|}\\
\beta_{j+1}(a_j) = \lim_{t\rightarrow a_j}\frac{\gamma_{j+1}(t)}{|\gamma_{j+1}(t)|} = -\frac{\dot{\gamma}(a_j)}{|\dot{\gamma}(a_j)|}\\
\end{split}
\end{equation} Let $\tilde{\beta}_j(a_j)=(\beta_j(a_j), \pm \hat{u})\in L(4,1)$ for some $\hat{u}$ normal to $\beta_j(a_j)$, following lemma 1. We may choose
\begin{equation}
\tilde{\beta}_{j+1}(a_{j})= (\beta_{j+1}(a_j), \pm \hat{u}) \in \pi_1^{-1}(\beta_{j+1}(a_j))
\end{equation}
This is a valid choice because $\hat{u}$ is normal to $\beta_{j+1}(a_j)$ (this follows from $\beta_{j+1}(a_j)= -\beta_j(a_j)$). It now follows that $\pi_2\circ \beta_{j}(a_j) = \pi_2 \circ \beta_{j+1}(a_j)= \pm \hat{u} \in \mathbb{RP}^2$ and therefore, $\tilde{\gamma}_j$ and $\tilde{\gamma}_{j+1}$ can be attached continuously .

It remains to show that the lift $\tilde{\gamma}$ obtained by attaching $\pi \circ \tilde{\gamma}_j$  minimizes the Fubini-Study metric. It suffices to show this for the interior of each segment $\pi \circ \tilde{\gamma}_j$, which is contained in $L(4,1)\times (0,1)$. Consider $\tilde{\gamma}_j(t)=(r(t), \hat{v}(t), \pm \hat{u}(t))$ as a path in the set of all chords to a unit sphere, following lemma 2(c) and using the notation $(r, \hat{v}, \pm \hat{u})$ for a chord with center at $r\hat{v}$ and in direction $\hat{u}$. It follows from the construction of $\tilde{\gamma}_j$ that:
\begin{equation}
\begin{split}
r(t) & = |\gamma_j(t)| \\
(\hat{v}(t), \pm \hat{u}(t))& = \tilde{\beta}_j(t) \in L(4,1) \quad \text{and}\\
\hat{v}(t) &= \beta_j(t) \\
\end{split}
\end{equation}
$r(t)$ and $\hat{v}(t)$ are determined by $|\gamma_j(t)|$ and $\beta_j(t)$ respectively. The key observation is that the horizontal lift $\tilde{\beta}_j$ minimizes the length under the induced round metric on $L(4,1)$ (\Eqref{Metric on L41}) among all lifts of $\beta_j$ \cite{Kobayashi}, \cite{Kobayashi2}. That is,  $\hat{u}(t)$ is chosen so as to minimize the length of $\tilde{\beta}_j$ in $L(4,1)$. From \Eqref{Metric on L41}, it follows that $\dot{\hat{u}}\cdot \dot{\hat{u}} = (\hat{v}\cdot \dot{\hat{u}})^2$ (here, $\dot{\hat{u}}=\frac{d\hat{u}}{dt}$). This is the condition for minimizing the length. From \Eqref{Fubini-Study}, it follows that the same condition minimizes the Fubini-Stuidy length of $\tilde{\gamma}_j$ in $L(4,1)\times(0,1)$. Thus, $\tilde{\gamma}$ is a horizontal lift of $\gamma$. $\blacksquare$

\begin{figure}[h!]
\includegraphics[scale=0.57]{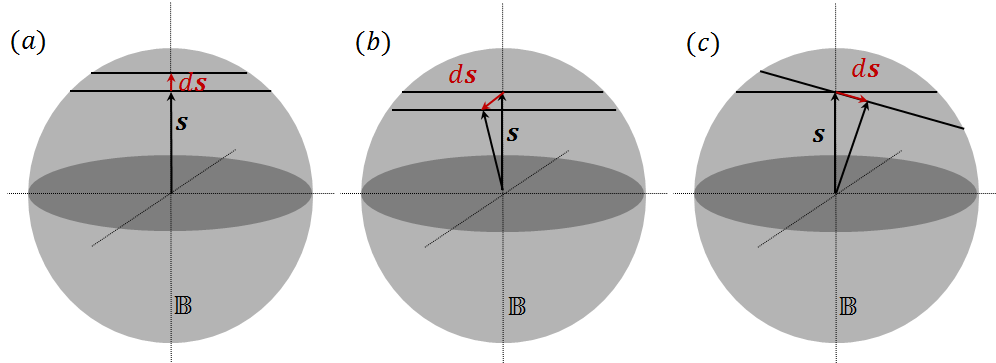}
\caption{\textbf{Parallel transport.} Horizontal lifts (i.e., parallel tranports) are defined by minimizing the Fubini-Study metric on $\mathbb{CP}^2$, which turns out to be equivalent to the intuitive notion of parallel transport similar to tangent vectors. Quantum state vectors can be represented by chords inside $\mathbb{B}$. The spin vector is given by the center of the chord. Corresponding to a change $d\vec{s}$ in the spin vector, the chords can be parallel transported. (a) shows the parallel transport of the chord when $d\vec{s}$ is parallel to $\vec{s}$. (b) shows parallel transport of the chord when $d\vec{s}$ is perpendicular to the chord and $\vec{s}$. (c) shows the parallel transport when $d\vec{s}$ is parallel to the chord.}\label{FIG5}
\end{figure}

We next demonstrate that the horizontal lift defined by minimizing the Fubini-Study metric is equivalent to the intuitive notion of parallel transport of ellipsoids inside $\mathbb{B}$. It is easier to use chords in $\mathbb{B}$, following lemma 2(c), instead of ellipsoids. Let $\psi$ be a quantum state vector represented by the chord $(r, \hat{v}, \pm\hat{u})$. Its spin vector is $\vec{s}=r\hat{v}$. We show that for every infinitesimal change $d\vec{s}$ of the spin vector, the corresponding intuitive parallel transport of the chord also minimizes the Fubini-Study metric. $d\vec{s}$ is a 3-dimensional vector and it can be written as a superposition of $\hat{v}$, $\hat{u}$ and $\hat{v}\times \hat{u}$. It suffices to consider the three cases where $d\vec{s}$ is parallel to the above mentioned three vectors separately. Let us begin with the case $d\vec{s}=|d\vec{s}|\hat{v}$ (\figref{FIG5} (a)). Intuitively, the chord should be moved radially, parallel to itself, i.e., after parallel transport, the new chord will be $(r+|d\vec{s}|, \hat{v}, \pm \hat{u})$. From \Eqref{Fubini-Study}, it follows that this is consistent with the minimization of the Fubini-Study metric. When $d\vec{s}$ is perpendicular to both $\hat{v}$ and $\hat{u}$, intuitively, the chord should be moved to $(r, \hat{v}+\frac{d\vec{s}}{r}, \pm \hat{u})$--- consistent with minimization of Fubini-Study metric (\figref{FIG5} (b)). Finally, when $d\vec{s}$ is parallel to $\hat{u}$, the chord should be parallel transported like a tangent line to the shell of radius $r$. Using straightforward geometry, it follows that the new chord is $(r, \hat{v}+\frac{d\vec{s}}{r}, \pm(\hat{u}-\frac{|d\vec{s}|}{r}\hat{v}))$ (\figref{FIG5} (c)). This satisfies the correct minimization condition for the Fubini-Study metric, $d\hat{u}\cdot d\hat{u}=(\hat{v}\cdot d\hat{u})^2$.

Geometric phase was defined in \secref{sec 2} as the operator $R\in SO(3)$ such that $\tilde{\gamma}(1)=\mathcal{D}(R)\tilde{\gamma}(0)$ holds for all lifts $\tilde{\gamma}$ of $\gamma$. However, this operator is not unique --- it has a two fold ambiguity because $\tilde{\gamma}(0)$ has a non-trivial stabilizer in $SO(3)$. For instance, when $|\vec{s}|\neq 0$, $R_{\vec{s}}(\pi)\tilde{\gamma}(0)=\tilde{\gamma}(0)$. We now use the details of the construction of $\tilde{\gamma}$ to clear this ambiguity and provide a rigorous definition of $R$.

Corresponding to each segment $\beta_j$ in $S^2$, we define a vertical displacement $R_j \in SO(3)$ such that its lift satisfies $\tilde{\beta}_j(a_j)= R_j \tilde{\beta}_j(a_{j-1})$. Here, $\tilde{\beta}_j$ is considered a path in the space of tangent lines to a sphere and $R_j$ acts on the tangent lines as a rotation. To define $R_j$ uniquely, we note that $SO(3)\approx L(2,1)$ is a double cover of $L(4,1)$. As remarked earlier, $L(2,1)$ is the space of all unit tangent vectors to a unit sphere. $\tilde{\beta}_j$ can be lifted to $L(2,1)$, and the end points of this lift will define a unique $R_j \in SO(3)$. For example, if $\tilde{\beta}_j(t) = (\hat{v}(t), \pm \hat{u}(t))$, we may assume without loss of generality, that $(\hat{v}(t), \hat{u}(t))$ represents a continuous path in the space of all unit tangent vectors, i.e., in $L(2,1)$. Indeed, this is a lift of $\tilde{\beta}_j$ in $L(2,1)$. The only other lift is $(\hat{v}(t), -\hat{u}(t))$. Both of these lifts define the same, unique vertical displacement $R_j\in SO(3)$ with
\begin{equation}
R_j \hat{v}(a_{j-1})= \hat{v}(a_j) \quad \text{and} \quad R_j \hat{u}(a_{j-1})= \hat{u}(a_j)
\end{equation}
Noting that $L(4,1)$ is a $U(1)$ bundle over $S^2$, it is straightforward to show that this operator is independent of the choice of the first point, $\tilde{\beta}_j(a_{j-1})$ of the lift \cite{Kobayashi}. We now define the geometric phase as
\begin{equation}\label{def of gp}
R= R_{n+1}R_n\cdots R_1
\end{equation}
It follows that $\mathcal{D}(R)\tilde{\gamma}(0)=\tilde{\gamma}(1)$.

We end this section with an explicit formula to compute the horizontal lift in $\mathbb{CP}^2$ and the geometric phase of a given loop in $\mathbb{B}$. It suffices to compute $\tilde{\beta}_j$ and $R_j$ for each piece $\gamma_j$ of the loop. Assuming that $\beta_j = \hat{v}(t)$ for $t\in [a_j, a_{j+1}]$, we are to find a $\hat{u}(t)$ such that $(\hat{v}(t), \pm \hat{u}(t))$ is a horizontal lift of $\beta_j$ with a given initial point $\hat{u}(0)$. Using the minimization condition for \Eqref{Metric on L41} and $\hat{u}(t)\cdot \hat{v}(t)=0$, it follows that $\hat{u}(t)$ is the solution to the differential equation with the given initial value:
\begin{equation}\label{eqn_u}
\frac{d}{dt}\hat{u}(t) = - \left(\frac{d\hat{v}(t)}{dt}\cdot \hat{u}(t)\right)\hat{v}(t)
\end{equation}
To find the geometric phase, we introduce $X:[a_j,a_{j+1}]\rightarrow SO(3)$ satisfying $\hat{u}(t)=X(t)\hat{u}(a_j)$, $\hat{v}(t)=X(t)\hat{v}(a_j)$ and $X(a_j)=1$. The geometric phase will then be $R_j= X(a_{j+1})$. It is straightforward to see that $X(t)$ is the solution to the following initial value problem:
\begin{equation}\label{eqn_R}
\begin{split}
&\frac{d}{dt}X(t)= \left(\frac{d\hat{v}(t)}{dt}\hat{v}(t)^T-\hat{v}(t)\frac{d\hat{v}(t)}{dt}^T\right)X\\
&X(a_j) = 1\\
\end{split}
\end{equation}
The above two equations, along with \Eqref{CP2-chord} provide a complete set of equations to compute the horizontal lift and the geometric phase for any loop in $\mathbb{B}$. 

Before proving theorem 2, we make a few remarks regarding the points on the boundary of $\mathbb{B}$. The pre-image set of these points is trivial (\Eqref{pre-image sets}). This implies that the corresponding quantum states can not carry any geometric phase information. Nevertheless, the definition of horizontal lifts and geometric phase given above are valid even for loops that visit the boundary of $\mathbb{B}$. 

To understand what the horizontals lift and geometric phases of the class of loops that visit the boundary of $\mathbb{B}$ mean, we note that such loops can be pushed to the interior of  $\mathbb{B}$ through infinitesimal perturbations. It is straightforward to see that the horizontal lift (geometric phase) of such loops is indeed equal to the limit of the horizontal lift (geometric phase) of the perturbed loops. Therefore, although no geometric phase can be extracted physically from this particular class of loops, for the purpose of theoretical completeness, it is possible to consistently define a geometric phase for them.

\subsection{Proof of Theorem 2}\label{sec 3b}

As shown in lemma 1, $L(4,1)$ admits two $S^1$ bundle structures, namely, $\pi_1: L(4,1)\rightarrow S^2$ and $\pi_2 : L(4,1)\rightarrow \mathbb{RP}^2$.  Accordingly, loops in $S^2$  and loops in $\mathbb{RP}^2$ both have well-defined solid angles in terms of the respective $U(1)$ holonomies. The natural projection from $S^2$ to $\mathbb{RP}^2$ preserves the solid angle.  This is the core ingredient in the interpretation of the geometric phase and the proof of theorem 2. We prove this fact in lemma 3 and then proceed to prove theorem 2. We denote the natural projection map from $S^2$ to $\mathbb{RP}^2$ by $p$.

\textbf{Lemma 3:} Let $\beta$ be a piece-wise differentiable path in $S^2$ and $p\circ \beta$ be its projection in $\mathbb{RP}^2$. The vertical displacements of the horizontal lifts of $\beta$ and $p\circ \beta$ in $L(4,1)$ are equal.

\textbf{Proof:} Let $\beta(t)=\hat{v}(t)$ and let $\tilde{\beta}(t)= (\hat{v}(t), \pm \hat{u}(t))$ be its horizontal lift in $L(4,1)$. The projection of $\beta$ in $\mathbb{RP}^2$ is $p\circ \beta = \pm \hat{v}(t)$. We first show that the path obtained by interchanging the two vectors $\hat{u}$ and $\hat{v}$ in $\tilde{\beta}$, i.e., $(\hat{u}(t), \pm \hat{v}(t))$, is a horizontal lift of $p\circ \beta$ in $L(4,1)$.

From the condition $\hat{u}(t)\cdot \hat{v}(t)=0$, it follows that $\dot{\hat{u}}(t)\cdot \hat{v}(t)+\hat{u}(t)\cdot \dot{\hat{v}}(t)=0$. Therefore, the paths $(\hat{v}(t), \pm \hat{u}(t))$ and $(\hat{u}(t), \pm \hat{v}(t))$ have the same length in $L(4,1)$(see \Eqref{Metric on L41}).  Further, $(\hat{u}(t), \pm \hat{v}(t))$ is a lift of $p\circ\beta$ because, $\pi_2\circ (\hat{u}(t), \pm \hat{v}(t))= \pm \hat{v}(t) = p\circ \beta(t)$. We show, by contradiction, that it is indeed a horizontal lift. If it is not a horizontal lift, let $(\hat{u}'(t), \pm \hat{v}(t))$ be the unique horizontal lift with the initial value $\hat{u}'(0)=\hat{u}(0)$. It must have a shorter length than $(\hat{u}(t), \pm \hat{v}(t))$. It follows now that $(\hat{v}(t), \pm \hat{u}'(t))$ is a lift of $\beta$ with a length shorter than $\tilde{\beta}(t)=(\hat{v}(t), \pm \hat{u}(t))$, and they have the same initial point i.e., $(\hat{v}(0), \pm \hat{u}'(0)) = (\hat{v}(0), \pm \hat{u}(0))$. This contradicts with the hypothesis that $\tilde{\beta}$ is a horizontal lift.

Thus, $\tilde{p\circ\beta}=(\hat{u}(t), \pm \hat{v}(t))$ is a horizontal lift of $p\circ \beta$. Let us now consider lifts of $\tilde{\beta}$ and $\tilde{p\circ\beta}$ in $L(2,1)$ i.e., $(\hat{v}(t), \hat{u}(t))$ and $(\hat{u}(t), \hat{v}(t))$ respectively. It is straightforward to see that the vertical displacements are identical and is given by
the unique $SO(3)$ operator $V$ which satisfies $V\hat{v}(0)=\hat{v}(1)$ and $V\hat{u}(0)=\hat{u}(1)$. $\blacksquare$

We now return to prove theorem 2. Although the pieces $\beta_j$ in $S^2$ cannot be attached continuously, their projections in $\mathbb{RP}^2$ can be attached continuously:
\begin{equation}
p\circ\beta_j(a_{j})= \pm \frac{\dot{\gamma}(a_j)}{|\dot{\gamma}(a_j)|}= p\circ\beta_{j+1}(a_{j})
\end{equation}
This follows from \Eqref{beta}. Indeed, the path obtained by attaching the segments $p\circ \beta_j$ in $\mathbb{RP}^2$ is $\alpha$, the projection of $\gamma$ defined in \Eqref{Projection to rp2}. From lemma 3, it follows that the vertical displacements of $\beta_j$ and $p\circ\beta_j$ are equal. Thus, the vertical displacement of $\alpha$ is given by
\begin{equation}
V=R_{n+1}R_n \cdots R_1
\end{equation}
Where, $R_j$ is the vertical displacement of $\beta_j$. This is equal to the geometric phase of $\gamma$, defined in \Eqref{def of gp}.$\blacksquare$

\subsection{Generalized Solid Angle}\label{sec 3c}
The notion of generalized solid angle was introduced through definition 3 in \secref{sec 2}. In the following, we show that this generalized solid angle reduces to the standard solid angle for non-singular loops. Furthermore, we discuss the reasons why this definition is a meaningful generalization of solid angles for singular loops. In particular, we discuss the case when the projected path $\alpha$ is open in $\mathbb{RP}^2$.

When $\gamma$ is non-singular, its projection $\alpha$ is necessarily closed. We consider the following three cases separately --- (i) $\gamma$ is non singular, (ii) $\gamma$ is singular and $\alpha$ is closed and (iii)  $\gamma$ is singular and $\alpha$ is an open path.

For a non-singular loop, by definition $|\gamma(t)|\neq 0$ throughout. Therefore it comprises of only one piece, i.e., $a_0=0$ and $a_1=1$. The corresponding projected path in $S^2$, $\beta = \frac{\gamma}{|\gamma|}$ is closed. From lemma 3 and the definition of the geometric phase given by \Eqref{def of gp}, it follows that the geometric phase ($R$) of $\gamma$ is a rotation about $\beta (0)$ (or equivalently, about $\alpha (0)$) by an angle equal to the solid angle of $\gamma$. This angle is obtained by the expression $\cos^{-1}(\hat{k}\cdot R\hat{k})$ for some unit vector $\hat{k}$ normal to $\alpha(0)$. Thus, the generalized solid angle is consistent with the standard solid angle for non-singular loops.

\begin{figure}[h!]
\includegraphics[scale=0.7]{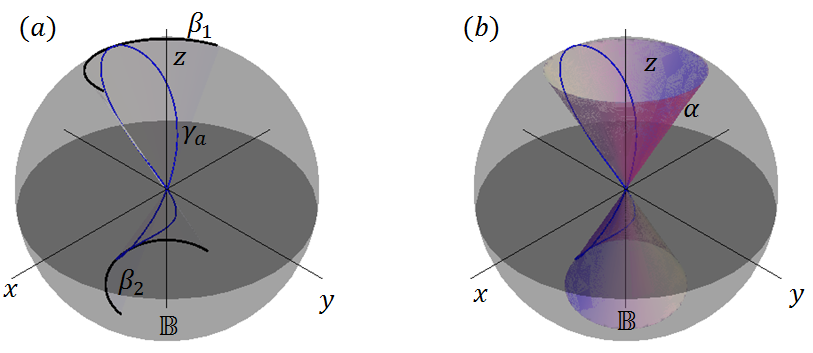}
\caption{\textbf{Generalized solid angle}: (a) shows a singular loop $\gamma_a$ (in blue) and the segments of its projection to $S^2$, $\beta_1$ and $\beta_2$ (in black). (b) shows the projection of $\gamma_a$ to $\mathbb{RP}^2$, $\alpha$. Every point in $\mathbb{RP}^2$ is a diamter of $S^2$ and therefore, $\alpha$ is obtained by mapping every point on $\gamma_a$ to the respective diameter of $\mathbb{B}$. The resulting cone is a loop in the space of diameters to $S^2$, i.e., in $\mathbb{RP}^2$. Thus, the cone represents $\alpha$ and the generalized soild angle of $\gamma_a$ is indeed equal to the soild angle of the cone.  }\label{FIG6}
\end{figure}
For a singular loop, the standard solid angle is not well-defined. However, if the projection $\alpha$ is closed, i.e., $\alpha(0)=\alpha(1)$, the geometric phase (i.e., the vertical displacement of $\alpha$) is still a rotation about $\alpha(0)$ --- it maps the fiber above $\alpha(0)$ in $L(4,1)$ to itself. Therefore, the angle of rotation about $\alpha(0)$ is well-defined and is the natural extension of solid angles to this case.

Finally, we consider the case where $\alpha$ is open. \figref{FIG3}(b) shows one such example of a loop $\gamma$, whose projection is open in $\mathbb{RP}^2$. That is, $\gamma(0)=\gamma(1)=\vec{0}$ but $\pm \frac{\dot{\gamma}(0)}{|\dot{\gamma}(0)|}=\alpha(0)\neq \alpha(1)=\pm \frac{\dot{\gamma}(1)}{|\dot{\gamma}(1)|}$. Solid angles are well-defined for open paths in $S^2$ by closing them using a geodesic in \cite{Pancharatnam1956}, \cite{PhysRevLett.60.2339} (see also ref. \cite{PhysRevA.52.2576} for an alternative formulation). We adopt a similar technique to define solid angles for open paths in $\mathbb{RP}^2$. The geometric phase ($R$) maps the fiber above $\alpha(0)$ to the fiber above $\alpha(1)$ in $L(4,1)$. Indeed, it can be written uniquely as a product of two rotations, one that takes $\alpha(0)$ to $\alpha(1)$ and another that rotates about $\alpha(1)$:
\begin{equation}\label{GSA}
R = R_{\alpha(1)}(\Omega_2)R_{\hat{k}}(\Omega_1)
\end{equation}
where $\hat{k}$ is a vector normal to $\alpha(0)$ and $\alpha(1)$ and $\Omega_1$ is the angle between $\alpha(0)$ and $\alpha(1)$. The natural definition of solid angle for such a path is $\Omega_2$, which is given by $\cos^{-1}(\hat{k}\cdot R\hat{k})$.

\section{Examples and Experimental Considerations}\label{sec 4}
In this section, we discuss two examples and make a few remarks on how this geometric phase can be observed experimentally.  To begin, we present the general procedure to determine the Horizontal lift and geometric phase for any given loop, in five steps:
\begin{itemize}
\item [1.] Verify if the loop is liftable. If it is not liftable, the horizontal lift and the geometric phase are not well defined.
\item[2.] Identify the ``zeros" of the loop, i.e., the set $\{a_0, a_1, \cdots , a_{n+1}\}= \gamma^{-1}(\vec{0})$ at which the loop visits the center of $\mathbb{B}$. Because the loop is liftable, it will be differentiable at each $a_i$.
\item[3.] Divide the loop into the segments $\gamma_i:[a_{i-1}, a_i]\rightarrow \mathbb{B}$ and determine their projections $\beta_i(t)$, as defined in the proof of theorem-1.
\item[4.] Solve\Eqref{eqn_u} for $\tilde{\beta}_i(t)$ (and hence obtain $\tilde{\gamma}_i(t)$) and \Eqref{eqn_R} for $X_i(t)$ and hence obtain $R_i$.
\item[5.] By concatenating $\tilde{\gamma}_i$, obtain the horizontal lift $\tilde{\gamma}$, i.e., the lift that minimizes the Fubini -Study length. The geometric phase is, using \eqref{def of gp}, $R=R_{n+1}R_n\cdots R_1$ and the generalized solid angle is given by $\cos^{-1}(\hat{}k\cdot R\hat{k})$ for some unit vector $\hat{k}$ normal to both $\beta_1(0)$ and $\beta_n(1)$. 
\end{itemize}

\begin{figure}[h!]
\includegraphics[scale=0.7]{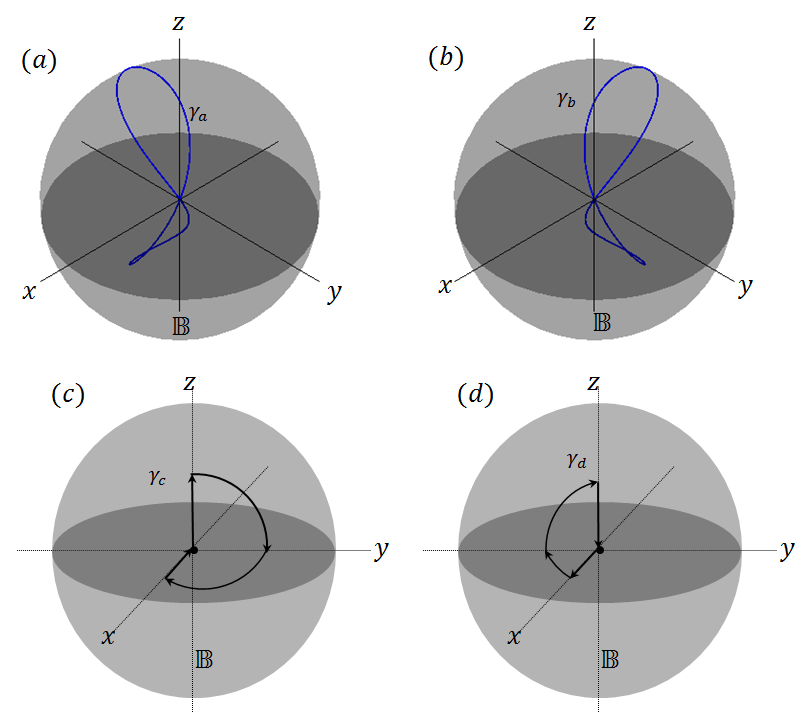}
\caption{\textbf{Liftable-singular loops}: All of the four loops shown start and end at the center. The product of the loops (a) and (b) is not liftable. (c) and (d) on the other hand have a liftable product and further, their geometric phasees are non-commuting. Thus, they illustrate the non-Abelian nature of this geometric phase.}\label{FIG7}
\end{figure}

We now illustrate this procedure to compute the generalized solid angle of the singular loop shown in \figref{FIG2}(b), as a first example. This is also the loop ``$a$" in \figref{FIG7}. In Cartesian coordinates, this loop is (using the notation $\gamma_a(t)=(x(t), y(t), z(t))$): 
\begin{equation}
\gamma_a(t)=(\sin (2\pi t)\sin (\pi/6)\cos(2\pi t), -\sin (2\pi t)\sin (\pi/6)\sin(2\pi t), \sin (2\pi t)\cos (\pi/6))
\end{equation} In spherical polar coordinates, this loop can be represented as $r(t)=\sin(2\pi t)$ (allowing negative values), $\theta(t)=\pi/6$ and $\phi(t)=-2\pi t$. We now follow the procedure outlined above to compute its horizontal lift, geometric phase and generalized solid angle:
\begin{itemize}
\item [1.] This loop is liftable, because $\gamma_a(t)$ is differentiable everywhere.
\item[2.] This loop has zeros at $t=0, 1/2 \ \&\ 1$, i.e, Thus, $n=1$ and $a_0=0, a_1=1/2, a_2=1$.
\item[3.] The segments $\gamma_1$ and $\gamma_2$ are given by restricting the loop to $[0,1/2]$ and $[1/2, 1]$ respectively. The corresponding projections $\beta_i$, spherical polar coordinates, are given by:
\begin{equation*}
\begin{split}
\beta_1(t) = &(\frac{1}{2} \cos(
2\pi t), -\frac{1}{2} \sin(2\pi t), \frac{\sqrt{3}}{2}), \ t\in (0,1/2)\\
\beta_2(t)=&(-\frac{1}{2} \cos(
2\pi t), \frac{1}{2} \sin(2\pi t),-\frac{\sqrt{3}}{2}),  \ t\in (1/2,1)\\
\end{split}
\end{equation*}
\figref{FIG6} (a) shows these projections.
\item[4.] \Eqref{eqn_u} and \Eqref{eqn_R} both have analytical solutions. In particular, the solutions to\Eqref{eqn_R} are:
\begin{equation*}
\begin{split}
X_1(t)&=R_{\beta_1(t)}(-\sqrt{3}\pi t)R_z(2\pi t): \ \  t \in [0, 1/2] \\ 
X_2(t)&=R_{\beta_2(t)}(\sqrt{3}\pi (t-1/2))R_z(2\pi (t-1/2)): \ \  t \in [1/2, 1] \\ 
\end{split}
\end{equation*}
This indeed implies that \Eqref{eqn_u} also has an analytic solution. Using the notation from the proof of theorem-1, $\tilde{\beta}_i(t)=(\beta_i(t), \pm u_i(t))$ for $i=1,2$, it is straightforward to see that $u_1(t)=X_1(t)u_1(0)$ and $u_2(t)=X_2(t)u_1(1/2)$. The horizontal lifts depend on the choice of the initial point, $u_1(0)$. 
\item[5.] Thus, $R_1=R_{\beta_1(1/2)}(-\sqrt{3}\pi /2)R_z(\pi)= R_z(\pi)R_{\beta_1(0)}(-\sqrt{3}\pi /2)$ and $R_2= R_{\beta_2(1)}(\sqrt{3}\pi/2)R_z(\pi)$. Noting that $R_z(2\pi)=R_{\beta_1(0)}(2\pi)$ and $\beta_1(0)=-\beta_2(1)$, it follows that the geometric phase is $R_a=R_2R_1 = R_{\beta_1(0)}((2-\sqrt{3})\pi)$. Following \Eqref{GSA}, the generalized solid angle is $\Omega =\pi (2-\sqrt{3})$. The most straightforward way to obtain the concatenated lift $\tilde{\gamma}$, starting from an initial point $\tilde{\gamma}(0)$ is using the equation $\tilde{\gamma}(t)= (|\gamma_a(t)|, \tilde{\beta}(t))$, where we are representing a quantum state by a chord.
\end{itemize}
The projection of this loop to $\mathbb{RP}^2$ is $\alpha(t)=\pm (\frac{1}{2} \cos(
2\pi t), -\frac{1}{2} \sin(2\pi t), \frac{\sqrt{3}}{2})$. If every point in $\mathbb{RP}^2$ is represented by a diameter of $S^2$, this loop represents a cone (see \figref{FIG6}(b)). The generalized solid angle of this loop is indeed equal to the solid angle of the cone, as shown by theorem 2. 

The geometric phase of this loop, $R_a=R_{\beta_1(0)}((2-\sqrt{3})\pi)$ is a rotation about the axis, $(1/2, 0, \sqrt{3}/2)$ by an angle $(2-\sqrt{3})\pi$. The axis is the tangent to the loop at $t=0$. This already indicates that geometric phase of two loops based at the center can be non-commuting. If, for instance, we choose another loop congruent to this one but oriented differently (see \figref{FIG7}(b)), its geometric phase will also be a rotation by an angle $(2-\sqrt{3})\pi$, but about a different axis because the loop would start with a different tangent. For instance, \figref{FIG7}(b) shows a loop $\gamma_b$, congruent to $\gamma_a$, obtaned by rotating the latter about the $z-$axis by $\pi/2$. Its geometric phase is $R_b=R_z(\pi/2)R_aR_z(-\pi/2)$. Although the two geometric phases do not commute, the product of the geometric phases is not equal to the geometric phase of the product of the loops, because the latter is not liftable. 

We provide another example, illustrating the non-Abelian nature of this geometric phase, where the geometric phases of two liftable loops do not commute and the concatenations of the loops are liftable.  These loops are also experimentally implemented easily, as we argue below. Let us consider the loops $\gamma_c$ and $\gamma_d$ shown in \figref{FIG7}(c) and \figref{FIG7}(d) respectively. It suffices to work out one of them in detail. $\gamma_c$ can be parametrized as:
\begin{equation*}
\gamma_c(t)=\begin{cases}
(0, 0, 2t):\ \ 0\leq t\leq 1/4 \\
(0, \frac{1}{2} \sin (2\pi(t-1/4)), \frac{1}{2} \cos (2\pi(t-1/4))): \ \ 1/4 \leq t\leq 1/2\\
( \frac{1}{2} \sin (2\pi(t-1/2)), \frac{1}{2} \cos (2\pi(t-1/2)), 0): \ \ 1/2 \leq t\leq 3/4\\
(2(1-t), 0, 0):\ \ 3/4\leq t\leq 1 \\
\end{cases}
\end{equation*}
We now follow the procedure to compute this loop's horizontal lift and geometric phase:
\begin{itemize}
\item[1.] Despite its appearance, this loop does \textit{not} have a non-differentiable point at the center; it visits zero only twice, at $t=0$ and $t=1$, under the given parametrization. And it is differentiable at both these times. Therefore, this loop \textit{is liftable}.
\item[2.] The set $\gamma_c^{-1}(\vec{0})$ is quite obviously $\{0,1\}$. 
\item[3.] There is only one segment, and its projection $\beta (t)$ is given by:
\begin{equation*}
\beta(t)=\begin{cases}
(0, 0, 1):\ \ 0\leq t\leq 1/4 \\
(0,  \sin (2\pi(t-1/4)), \cos (2\pi(t-1/4))): \ \ 1/4 \leq t\leq 1/2\\
(  \sin (2\pi(t-1/2)),  \cos (2\pi(t-1/2)), 0): \ \ 1/2 \leq t\leq 3/4\\
(1, 0, 0):\ \ 3/4\leq t\leq 1 \\
\end{cases}
\end{equation*} 
\item[4.] The solution to \Eqref{eqn_R} are:
\begin{equation*}
X(t)=\begin{cases}
1:\ \ 0\leq t\leq 1/4 \\
R_x(-2\pi(t-1/4)): \ \ 1/4 \leq t\leq 1/2\\
R_z(-2\pi(t-1/2))R_x(-\pi/2): \ \ 1/2 \leq t\leq 3/4\\
R_z(-\pi/2)R_x(-\pi/2):\ \ 3/4\leq t\leq 1 \\
\end{cases}
\end{equation*}
In the axis-angle representation. The lift $\tilde{\beta}$ is easy to obtain using this solution. 
\item[5.] The geometric phase is $R_c=R_z(-\pi/2)R_x(-\pi/2)$. The lift can be obtained, again, by using the equation $\tilde{\gamma}(t)= (|\gamma_c(t)|, \tilde{\beta}(t))$.
\end{itemize}
Similarly, for $\gamma_d$, the geometric phase is $R_d = R_x(\pi/2)R_z(-\pi/2)$. Thus, their geometric phases do not commute. They both have a generalized solid angle of $\pi/2$. 

In order to experimentally observe this geometric phase, the physical spin -1 system must satisfy the following criteria: (i) the quantum state vector of the system must be controllable and, (ii) a complete state tomography, i.e., a measurement of all the components of the spin vector and the spin fluctuation tensor must be possible.  Physical systems that satisfy both of these criteria include trapped atoms, ions, nitrogen-vacancy centers, superconducting qubits and experiments using nuclear-magnetic resonance. In the following, we briefly describe how the loops shown in \figref{FIG7} can be induced and their geometric phase can be measured using trapped atoms. We consider a Bose-Einstein condensate of trapped rubidium atoms, which is a spin-1 system. The spin eigenstates corresponding to eigenvalues $0$ and $\pm 1$ are magnetically sensitive. Therefore, using a strong spatial gradient in the magnetic field, a condensate can be spatially separated into three clouds, corresponding to the eigenstates $0$ and $\pm 1$. This is also known as Stern-Gerlach separation. The population of atoms in each cloud  can be estimated by measuring the intensity of fluorescent light emitted by each cloud. These populations are proportional to $|z_0|^2$ and $|z_{\pm}|^2$ where $(z_{-1}, z_0, z_{+1})$ is the quantum state vector of the system. The elements of the spin vector and the spin fluctuation tensor can be extracted from such measurements done in different basis. This technique has been demonstrated in \cite{2012NatPhys305H}. Further, the loops shown in \figref{FIG7} can be induced in this system using rotating magnetic fields and microwave pulses. The curved parts of this loop can be induced by rotating the spin vector using a radio-frequency magnetic field and the linear parts can be induced using microwave transitions. These techniques have also been demonstrated recently \cite{Hoang23082016}. The geometric phase of these loops can be measured by starting with two condensates prepared in the same quantum state and inducing the loop on one of them and then comparing the measured spin fluctuation tensor for both of them. For instance, the ellipsoid at the center of the Bloch sphere degenerates into a disk and has two independent parameters. Therefore, in an experiment where a loop that starts and ends at the center is induced, geometric phase can be observed by measuring this pair of parameters of the spin fluctuation tensor. Such an experiment has been recently done in a system of ultracold Rubidium atoms \cite{Our_Expt}. 

\section{Conclusions}
To conclude, we have shown that the geometrical properties of a loop traversed by the spin vector inside the Bloch ball can be extracted from the spin fluctuation tensor of a spin-1 quantum state. This property crucially depends on the Fubini-Study metric on $\mathbb{CP}^2$, and it reflects the deep synchrony between the geometry of real space and the geometry of the abstract space of quantum states.

For a loop inside the Bloch ball we call a path in the space of quantum states, a lift, if it projects down to the given loop upon evaluating the expectation value of the spin vector at each point and a horizontal lift, if it also minimizes its length under the Fubini-Study metric. Not every loop inside the Bloch ball has a well-defined horizontal lift. A loop is liftable, i.e., has a well-defined horizontal lift if it has at least one parameterization under which it is differentiable at every visit to the center. We have defined a geometric phase corresponding to each loop in a Bloch ball in the form of an $SO(3)$ operator.

Among liftable loops,  those that actually visit the center, which we call singular loops, are particularly interesting because of two non-trivial properties --- first, their geometric phase is non-Abelian and second, they don’t have a well-defined solid angle and therefore prompt a generalization of the same. We have introduced the notion of generalized solid angles, defined for both singular and non-singular loops, which reduces to the standard solid angle for non-singular loops, and use it to provide an interpretation of the geometric phase of liftable-singular loops.

On the experimental side, we have discussed the possibility of observing this geometric phase in a system of a Bose-Einstein condensate of rubidium atoms. We have argued that in this system, it is possible to use the existing experimental techniques to induce the loops shown in \figref{FIG7} and the geometric phase can be observed by comparing the tomographies performed on the quantum state before and after inducing the loop. 

Although we have considered a spin-1 system, our analysis can be generalized to any spin system. A spin-$S$ system has independent moment tensors up to order $2S$. A natural extension of our work is to explore the geometric information carried by these higher order tensors.

We end with a remark on the possible theoretical applications of our geometric phase. One of the recent theoretical applications has been in characterizing topological phases of matter. Berry's phase along a loop in the parameter space of a Hamiltonian is given by the integral of the Berry curvature evaluated over the region enclosed by the loop \cite{chruscinski2004}. The total integral of the Berry curvature over the entire parameter space (usually the momentum space in condensed matter systems) is a topological invariant of the parameter space known as the \textit{Chern number}. A topological phase transition is characterized by a ``sudden change" of the the Chern number. Recent explorations \cite{KitaevChain}, \cite{PhysRevLett.112.130401}, have shown that mixed state generalizations of Berry's phase \cite{Uhlmann.1986}, \cite{PhysRevLett.85.2845} can also be used to characterize topological phase transitions.  The geometric phase introduced in this paper could also be used to characterize topological states of 1-dimensional quantum systems. In particular the nature of singular loops is akin to critical points of quantum phase transitions, an example of which can be seen in \cite{KitaevChain}.

\section{Acknowledgments}
We thank T. A. B. Kennedy for discussions and providing extensive feedback during the preparation of the manuscript. We thank M. S. Chapman, Carlos Sa de Melo and Matthew Boguslawski for stimulating insights and discussions. We also thank John Etnyre for help with the formulation.  Finally, we acknowledge support from the National Science Foundation, grant no. NSF PHYS-1506294.

\section*{Appendix A: Eigenvalues of the spin fluctuation tensor}
In this section, we derive the expressions for the quantum state vector with a given spin vector and the eigenvalues of its spin fluctuation tensor. The system has an $SO(3)$ symmetry, i.e., if $R\in SO(3)$ and $\mathcal{D}(R)\in SU(3)$ is its representation, under the transformation $\psi \rightarrow \mathcal{D}(R)\psi$, the spin vector transforms as $\vec{s}\rightarrow R\vec{s}$ and the spin fluctuation tensor transforms as $\mathbf{T}\rightarrow R\mathbf{T}R^T $.  Therefore, for the purpose of derivation of the eigenvalues of $\mathbf{T}$, without loss of generality we may assume that $\vec{s}=(0,0,|\vec{s}|)^T$. Any normalized quantum state vector $\psi=(z_{-1}, z_0, z_{+1})^T$, (i.e., $\langle \psi, \psi \rangle=1$) with this spin vector must satisfy:
\begin{equation}
\begin{split}
z_{-1}z_0^*+z_0z_{+1}^* = 0 \\
|z_{+1}|^2-|z_{-1}|^2 = |\vec{s}|\\
\end{split}
\end{equation}
 This follows from \Eqref{the map phi}. When $0<|\vec{s}|<1$, the solutions to the above equation are
 \begin{equation}
 \psi = \left( \begin{array}{c}
\sqrt{\frac{1-|\vec{s}|}{2}} e^{-i\theta}\\
0\\
\sqrt{\frac{1+|\vec{s}|}{2}} e^{i\theta}\\
 \end{array}
 \right)
 \end{equation}
Each $\theta \in [0,\pi)$ produces a distinct quantum state with spin vector equal to $\vec{s}$. From \Eqref{tensor}, the corresponding spin fluctuation tensor can me computed:
\begin{equation}
\mathbf{T}= \left(\begin{array}{ccc}
\frac{1}{2}+\frac{\sqrt{1-|\vec{s}|^2}}{2}\cos 2\theta & \frac{\sqrt{1-|\vec{s}|^2}}{2}\sin 2\theta & 0\\
\frac{\sqrt{1-|\vec{s}|^2}}{2}\sin 2\theta & \frac{1}{2}-\frac{\sqrt{1-|\vec{s}|^2}}{2}\cos 2\theta & 0\\
0 &0 & 1-|\vec{s}|^2\\
\end{array}
\right)
\end{equation}
One of the eigenvectors of this matrix is $\vec{s}=(0, 0, |\vec{s}|)^T$ with an eigenvalue $1-|\vec{s}|^2$. The other two eigenvalues are easily seen to be $\frac{1\pm \sqrt{1-|\vec{s}|^2}}{2}$.

\section*{Appendix B: The connection form of this geometric phase}
The well known parallel transport of tangent vectors on a sphere is formulated using the affine connection. However, the geometric phase introduced in this paper can not be formulated using affine connection because its features are compatible only with Ehresmann connection. In order to put our geometric phase in perspective with the other well known examples and to bridge the gap between affine connection and Ehresmann connection, in this section, we show how the former naturally generalizes to the latter and that minimizing the length is a concise way of defining parallel transports. 

We begin with an overview of affine connection. Let $M$ be an $n-$manifold and $TM$ be its tangent bundle. At a point $q\in M$, with local coordinates $(x^1,x^2, \cdots , x^n)$, the tangent plane, $T_qM$ is an $n$ dimensional vector space spanned by $e_{\mu}= \frac{\partial}{\partial x^{\mu}} $, $\mu = 1,2,\cdots, n$. Central to an affine connection is the covariant derivative, which comes from ``differentiating" the basis vectors, $\frac{\partial}{\partial x^{\nu}}e_{\mu}=\Gamma^{\sigma}_{\mu \nu}e_{\sigma}$. In the more formal language, the covariant derivative, denoted by ``$D$" is defined as $De_{\mu} = \Gamma^{\sigma}_{\mu \nu}e_{\sigma}\otimes dx^{\nu}$. 

The basic problem, that the affine connection is designed to solve is to define a parallel transport $\tilde{\gamma}$, i.e., a horizontal lift, in $TM$ for a path $\gamma$ in $M$. This can be restated as follows: for a point $q\in M$ on $\gamma$ with a given a lift $(q, v)\in TM $ on $\tilde{\gamma}$ (where $v=v^{\mu}e_{\mu} \in T_qM$) and the local tangent vector $y=y^{\mu}e_{\mu}$ of $\gamma$, along which $q$ is moved, how do we change the coordinates $v^{\mu}$ in order to maintain the vector $v$ parallel to itself? In other words, we are to decide the local tangent vector $t$ of $\tilde{\gamma}$, that moves the point $(q,v)\in TM$ such that $q$ is moved along $y$ and $v$ remains parallel to itself. This tangent vector is in the $2n$ dimensional tangent plane of $TM$ at $(p,v)$, i.e., $t\in T_{(q,v)}(TM)$ . This space is spanned by the basis vectors $(e_1, e_2, \cdots, e_n, f_1, f_2, \cdots , f_n)$, where $f_{\mu}=\frac{\partial}{\partial v^{\mu}}$. Quite obviously, $t=y^{\mu}e_{\mu}+z^{\nu}f_{\nu}$, for a suitable choice of coefficients $z^{\nu}$ such that $tDv=0$. This condition is the parallel transport criterion. Using $Dv= D(v^{\mu}e_{\mu})=dv^{\mu} e_{\mu}+ v^{\mu} De_{\mu}$, we get $z^{\mu}e_{\mu}+ \Gamma^{\sigma}_{\mu \nu}e_{\sigma} y^{\nu}v^{\mu} =0$. For convenience, the connection matrix is defined as $\omega_{\nu}^{\mu}=\Gamma^{\mu}_{\nu \sigma}y^{\sigma}$, in terms of which, $z^{\mu}=-\omega^{\mu}_{\nu}y^{\nu}$ \cite{Chern}.

In order to move towards Ehresmann connection, we rewrite the affine connection in a coordinate-free form. Two observations are crucial. First, in the $2n$ dimensional tangent space $T_{q, v}(TM)$, the $n$ dimensional subspace spanned by $f_{\mu}$ represent changes to the tangent $v$ alone and is therefore known as the \textit{vertical subspace} of $T_{(q,v)}(TM)$.  And second, the special vectors $t\in T_{(q,v)}(TM)$ that satisfy $tDv=0$ also form an $n$ dimensional subspace spanned by $\{e_{\mu}-\omega_{\mu}^{\nu}f_{\nu}: \ \mu= 1, 2, \cdots, n\}$. This space complements the vertical subspace and is known as \textit{horizontal subspace} of $T_{(q,v)}(TM)$.  It is straightforward to see that all that the affine connection does is to identify this horizontal subspace at each point of $TM$. Indeed, any $n$ dimensional subspace of $T_{(q,v)}(TM)$ that complements the vertical subspace uniquely defines the elements $\omega_{\mu}^{\nu}$ of the connection matrix. This follows from the fact that the vector $e_{\mu}$ can be written uniquely as a sum of two vectors, one in the horizontal subspace and one in the vertical subspace. Thus, in the coordinate-free form, a connection is a specification of a horizontal subspace at each point of $TM$, that complements the vertical subspace. Indeed, it is an $n$ dimensional distribution over $TM$.

This definition extends to any fiber bundle and is known as Ehresmann connection. Apart from being coordinate-free, it also has two other advantages over the affine and the Levi-Civita connections. The covariant derivative, uipon which the affine and Levi-Civita connections are based, is rooted in differentiating vector fields over a manifold (i.e., sections of the vector bundle). In a fiber bundle where the fiber has non-trivial topology,  in general there are no sections. The Ehresmann connection is therefore the most natural way of defining horizontal lifts. Furthermore, the Ehresmann connection generalizes naturally to structures that are not bundles. The dimension of the horizontal space can be non-uniform. Such a connection, however, can not be traced back to a connection matrix and therefore, regarding a connection as a distribution becomes inevitable. 

If a fiber bundle has a natural Riemannian metric, the horizontal subspace can be defined as the orthogonal complement of the vertical subspace. The resulting horizontal lifts mimimize the path length. The tangent vector to the horizontal lift is always confined to the local horizontal subspace and because the vertical subspace is orthogonal to it, any added component would only make the lift longer. We illustrate with two examples, that by minimizing the length of a lift, we recover the standard connection form:

\begin{itemize}
\item \textit{Berry phase for spin-1/2 systems:} Let $\hat{n}(t)$ be a loop on the Bloch sphere and $\psi(t)$ be one of its lifts in $\mathbb{C}^2$. We assume that $\psi(t)$ is normalized. All other lifts of $\hat{n}(t)$ with the same initial point are of the form $e^{ix(t)}\psi(t)$, where $x(t)$ is a real scalar with $x(0)=0$. The length of these lifts ($s$) is given, under the Euclidean metric, by $s=\int \sqrt{||\dot{\psi}(t)||^2 +|\dot{x}|^2+ i\dot{x}(\langle \dot{\psi}, \psi\rangle - \langle \psi, \dot{\psi} \rangle)} dt $. The first order functional derivative of $s$ w.r.t to $x(t)$ is $\langle \dot{\psi}, \psi\rangle - \langle \psi, \dot{\psi} \rangle$ and it should vanish if $\psi$ is a horizontal lift, according to the new definition. Together, with $||\psi(t)||=1$, we obtain the correct parallel transport criterion: $\langle \dot{\psi}, \psi\rangle=0$, i.e., the Berry connection form \cite{PhysRevLett.51.2167}. The same argument applies for Aharonov-Anandan phase \cite{PhysRevLett.58.1593}.

\item \textit{Wilczek-Zee phase \cite{PhysRevLett.52.2111}:} We follow the generalization of Wilczek-Zee phase in \cite{ANANDAN1988171}, and show that the connection form thereof can be derived by minimizing the length. Let us assume that the Hilbert space can be decomposed into a direct sum of two subspaces with dimensions $n$ and $m$ respectively (denoted by $V_n$ and $V_m$). $V_n$ is the time dependent $n$ dimensional eigenspace of the Hamiltonian. We are to define parallel transport of a vector $\psi\in V_n(0)$ in time, i.e., we are to define a path $\psi(t)\in V_n(t)$ with $\psi(0)=\psi$ and minimal length. Given any lift, $\psi(t)$, we can construct other such lifts by  a transformation $U(t)\psi(t)$, where $U(t)$ is a unitary operator with all of its non-trivial eigenvectors in $V_n(t)$. The length of such a path is $s=\int \sqrt{||\dot{\psi}(t)||^2 + \langle \dot{U}\psi|\dot{U}^{\dagger}\psi\rangle + \langle U^{\dagger}\dot{U}\psi|\dot{\psi}\rangle + \langle \dot{\psi} |U^{\dagger}\dot{U}\psi\rangle} dt $. Again, setting $U^{\dagger}\dot{U}=iH(t)$ for some Hermitian operator $H(t)$, the first derivative of $s$ is $\langle H(t)\dot{\psi}|\psi\rangle - \langle \psi |H(t)\dot{\psi} \rangle$. $H(t)$ is an arbitrary Hermitian with zero eigenvalues in $V_m(t)$. Therefore, $s$ is stationary iff $\dot{\psi}$ is orthogonal to $V_n(t)$, indeed the same condition as equation (7) in ref. \cite{ANANDAN1988171}. 
\end{itemize}

In our system, at most of the points in $\mathbb{CP}^2$, the vertical subspace is one dimensional and the horizontal subspace, defined by the Fubini-Study metric is three dimensional. However, at some points, i.e., within the pre-image of the center of the Bloch ball, the vertical and the horizontal subspaces are both two dimensional. This feature adds all the non-trivialities to the system and therefore, we have to use Ehresmann connection in this problem.

\bibliography{GeometricPhasePaper}{}
\bibliographystyle{ieeetr}
\end{document}